\documentclass{article}

\usepackage[utf8]{inputenc}
\usepackage{authblk}
\usepackage[letterpaper, margin=1in]{geometry}
\usepackage[square,sort&compress,numbers]{natbib}
\usepackage[version=3]{mhchem} 

\usepackage{graphicx}

\usepackage[table,xcdraw,dvipsnames]{xcolor} 

\usepackage{xr}
\usepackage{xr-hyper}
\usepackage{hyperref}
\hypersetup{
    colorlinks=true,
    linkcolor=blue,
    citecolor=blue,
    filecolor=blue,
    urlcolor=blue,
}

\makeatletter

\usepackage{setspace}

\title{Exploring DFT$+U$ parameter space with a Bayesian calibration assisted by Markov chain Monte Carlo sampling}
\author[1,*]{Pedram Tavadze}
\author[1]{Reese Boucher}
\author[1]{Guillermo Avenda{\~n}o-Franco}
\author[2]{Keenan X. Kocan}
\author[3]{Sobhit Singh}
\author[1]{Viviana Dovale-Farelo}
\author[4]{Wilfredo Ibarra-Hern\'andez}
\author[1]{Matthew B Johnson}
\author[2]{David S. Mebane}
\author[1]{Aldo H Romero}

\affil[1]{Department of Physics and Astronomy, West Virginia University, Morgantown, WV, USA}
\affil[2]{Department of Mechanical and Aerospace Engineering, West Virginia University, Morgantown, WV, USA}
\affil[3]{Department of Physics and Astronomy, Rutgers University, Piscataway, NJ, USA}
\affil[4]{Facultad de Ingenier\'ia, Benem\'erita Universidad Aut\'onoma de Puebla, Apdo. Postal J-39, Puebla, Pue. 72570, M\'exico }
\affil[*]{Corresponding author: petavazohi@mix.wvu.edu}

\date{January 2021}
\begin{document}

\maketitle

\begin{abstract} \label{abstract}
    Density-functional theory is widely used to predict the physical properties of materials. However, it usually fails for strongly correlated materials. A popular solution is to use the Hubbard corrections to treat strongly correlated electronic states. Unfortunately, the exact values of the Hubbard $U$ and $J$ parameters are initially unknown, and they can vary from one material to another. In this semi-empirical study, we explore the $U$ and $J$ parameter space of a group of iron-based compounds to simultaneously improve the prediction of physical properties (volume, magnetic moment, and bandgap). We used a Bayesian calibration assisted by Markov chain Monte Carlo sampling for three different exchange-correlation functionals (LDA, PBE, and PBEsol). We found that LDA requires the largest $U$ correction. PBE has the smallest standard deviation and its $U$ and $J$ parameters are the most transferable to other iron-based compounds. Lastly, PBE predicts lattice parameters reasonably well without the Hubbard correction.
\end{abstract}

\includegraphics[width=0.9\textwidth]{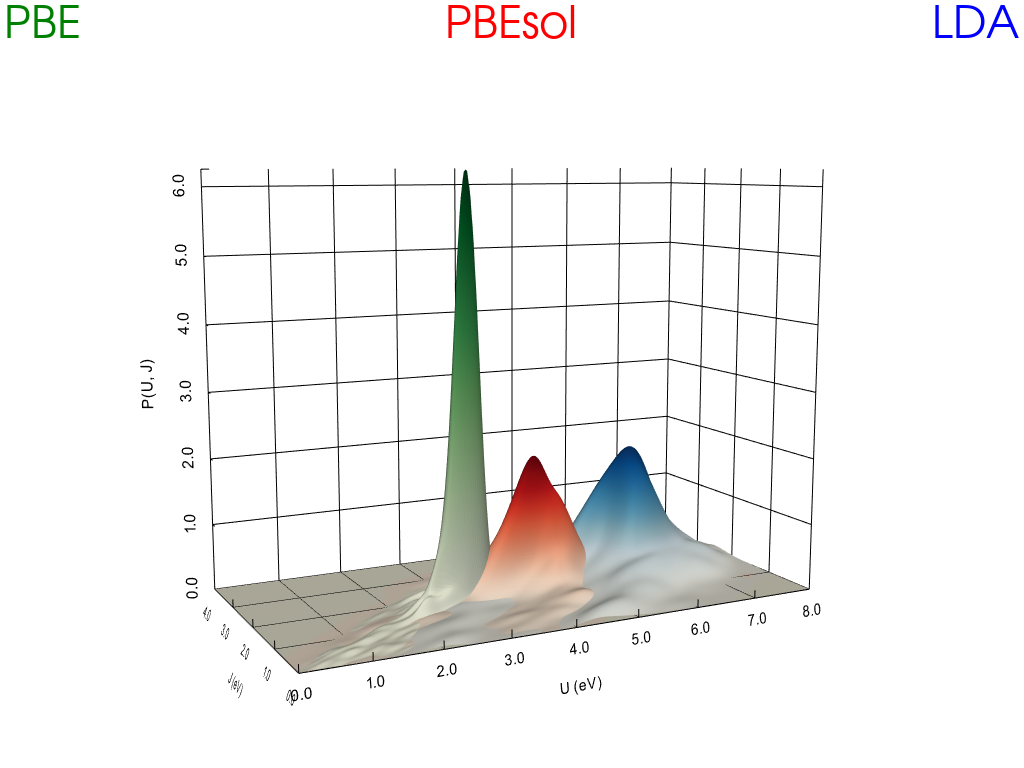}

\section{Introduction} \label{sec:intro}
Thanks to the seminal works of Hohenberg, Kohn, and Sham~\cite{hehenberg-kohn,kohn1965self,kohn1996density} researchers can simplify the many-body Schr{\"o}dinger's equation into a mean-field approach for the electronic Hamiltonian in materials. This approach allows us to computationally predict numerous material-specific properties utilizing the elegance of the density-functional theory (DFT)~\cite{kohn1965self,sholl2011density-DFT,kohn1996density,fiolhais2003primer-DFT,parr1980density-DFT}. Since the groundbreaking development of DFT, there have been numerous adaptations designed to optimize the accuracy of the exchange and correlation effects in DFT calculations. The largest complication of DFT lies within the accurate description of the exchange and correlation energy. An exact exchange-correlation (XC) functional is not yet known. However, various approximations for the XC functional have been made to more precisely and efficiently describe the electronic quantum states in materials~\cite{b88,pw91,lyp,PBE-PRB,PBEsol,SCAN-sun2015strongly,B3LYP1,MN15-L,XC-Progress,DFTthirtyyears}

Strongly correlated materials are greatly affected by the systematic error introduced in the widely used existing XC functionals, where the electronic kinetic energy is of the same order as the electron-electron repulsion. 
In this strong-interaction regime, distinct electronic properties can have various competing phases that are very sensitive to the description of the correlated-electronic states, as in the case of the $d-$ and $f-$electron systems, and in the metal-to-insulator transition observed in many oxides~\cite{MISRA2012199}. 
The lack of accurate representation of the electronic state by commonly used XC functionals impacts the prediction of the electronic and vibrational properties, 
in particular, the electronic bandgap, which can be significantly underestimated~\cite{2005-Heyd-EnergyBand,2017-Verma-HLE16}.

The currently accepted approaches to improve the DFT predictions, known as beyond-DFT methods, include: hybrid XC functionals~\cite{arbuznikov_hybrid_2007,perdew_hybrid,becke_hybrid,heyd_hybrid}, DFT+DMFT~\cite{DMFT-georges-kotliar1,DMFT-georges-kotliar-prb,kotliar2004strongly,kotliar2006electronic,georges2004strongly,Vollhardt_2011,DFTwDMFT_uthpala,dft_dmft1,kent2018dft_dmft, haule2015dft_dmft1, haule2015dft_dmft2, koccer2020dft_dmft, aichhorn2016dft_dmft, vollhardt2019dft_dmft}, and paramount to this work, DFT+$U$~\cite{DFT+U_LDA+U,DFT+U}. To address the above problem, DFT+$U$ introduces an on-site Coulombic interaction for the treatment of the electronic correlation effects~\cite{MISRA2012199}. An external Hubbard-like~\cite{1963-Hubbard,1964-Hubbard2} term is added to the DFT Hamiltonian along with a double-counting term, which negates the initial DFT calculation for the terms the Hubbard Hamiltonian attempts to correct. Two parameters $U$ and $J$ are supplemented to the Hubbard-like term to correct the Coulomb-repulsion term and the effective exchange interaction, respectively. This method is famously used in LDA+$U$~\cite{anisimov_dftu1,DFT+U_LDA+U,anisimov1997first_dftu2}, and can be generalized to numerous DFT functionals to correct the error-prone calculations. 

The main challenge facing DFT+$U$ is obtaining the optimal $U$ and $J$ correction parameters. To date, there are many methods designed to obtain these values. One of the most popular methods is the semi-empirical approach~\cite{wang2006oxidation_fitU} in which the parameters' values are modified until the DFT+$U$ predictions of some physical predefined observables are in agreement with the experimental measurements, such as electron bandgap, lattice parameters, or the atomic magnetic moment. Unfortunately, this method is limited to the materials for which experimental data is available.

Other methods are based on density-functional perturbation theory, linear response, the constrained random-phase approximation, and Hartree-Fock-based methods~\cite{pickett1998reformulation, Cococcioni_PhysRevB.71.035105,constrained_RPA_PhysRevB.74.125106,timrov2018hubbard, timrov2021self,csacsiouglu2011effective,vaugier2012hubbard, nakamura2021respack, mosey2007ab,agapito2015reformulation}. Though these theoretical methods are quite mature and have been implemented in different computational packages~\cite{QE_linear_response1,QE_linear_response2,segall2002CASTEP}, 
it is unclear if the search for optimal correctional parameters will have a unique global minimum, or multiple different local minima. This is a question that can only be addressed by a careful exploration of the $U$ and $J$ parameters. Furthermore, the explicit dependence of the DFT+$U$ Hamiltonian on orbital-dependence adds another dimension to the parameter space (\textit{i.e.}, the known metastability issue in DFT+$U$)  ~\cite{MeredigPRB2010, AllenWatson2014,payne2019PCCP}.

It is also unclear if a set of parameters defined for a specific material can be generalized to other materials (even within the same material family), or if the dependence of those parameters is strongly dependent on the selected XC functional within the DFT. The current understanding is that the correction parameters cannot be transferred to different materials because electronic correlations are governed by the nature of the chemical bonding and the coordination number, leading to the manifestation of different correlation effects within the same material family~\cite{kulik2010_bonding_dft,kulik2015_bonding_dftu,lany_zunger_2008_bonding_dft}. This further complicates the use of the DFT+$U$ methods in high-throughput calculations. 

In this investigation, we implemented an algorithm that builds a probability distribution in the parameter space of $U$ and $J$ for five strongly correlated iron-based compounds having different Fe oxidations states using three different XC functionals. We subsequently performed DFT+$U$ calculations using the mean values obtained for the $U$ and $J$ parameters for the initial five materials and three other similar iron-based compounds. We compared our results with the experimental data to investigate how well the distribution of the correction parameters can be extended to other similar compounds. Moreover, we inspected the relationship of the $U$ and $J$ parameters with different XC functionals.

\subsection{Bayesian Calibration and Markov Chain Monte Carlo Sampling}

The main goal of this project is to determine the distribution of the $U$ and $J$ values that can generate accurate predictions for iron-based materials using DFT$+U$ modeling. 
We use Bayesian calibration assisted by Markov chain Monte Carlo (MCMC) to sample the parameter space of $U$ and $J$ values on the potential energy surface. MCMC obtains the posterior distribution from the Bayes' theorem in an empirical form. 

Bayes' theorem defines the relationship between posterior and prior probability distributions on the parameter space:

\begin{equation}
    P({U,J}\mid X)=\frac{P(X\mid{U,J})~P({U,J})}{P(X)}.
\end{equation}

Where $P({U,J}\mid X)$ is the posterior density on the parameter space given the dataset $X$, $P(X\mid{U, J})$ is the likelihood, $P({U,J})$ is the prior density, and $P(X)=\int P(X\mid{U,J})P({U,J})dUdJ$ is the integrated probability of the data (or ``evidence") given the model.

\paragraph{Priors} The prior density is bounded uniform, with boundaries drawn in such a way that prevents the unphysical regions of the parameter space (\textit{i.e.}, $J > U$) from appearing in the posterior.

\paragraph{Likelihood} The likelihood model is a ``white noise" model with variance estimated in the course of the calibration 

\begin{equation}
    P(X \mid {U, J}) = \prod_j\frac{1}{(2\mathrm{\pi} \sigma_j)^{N_j/2}}\exp{\Bigl\lbrace\frac{\sum_i^{N_j} [M_{ij}(U, J) - X_{ij}]^2}{2\sigma_j^2}\Bigr\rbrace},
\end{equation}
where $M_{ij}$ and $X_{ij}$ are DFT model result and corresponding experimental measurement $i$ of type $j$, respectively, and $N_j$ is the total number of experimental results of type $j$. The variance of the experimental error $\sigma_j$ for property $j$ is estimated in the calibration, with an inverse gamma prior.

\paragraph{Markov chain Monte Carlo} The evidence $P(X)$ may be written in terms of the likelihood and prior 

\begin{equation}
    P(X) = \int P(X \mid {U, J})P({U, J})dJdU.
\end{equation}

This integral is not analytically estimable in the present case because of the nonlinear nature of the likelihood. Therefore, a Markov chain sampling procedure is used, which is guaranteed to converge in the limit of infinite samples drawn~\cite{1990-Gelfand-Sampling}.
In practice, the routine generally moves through an initial equilibration (burn-in) period before settling into its equilibrium state. 
Convergence is not guaranteed if insufficient samples are drawn from the parameter space, but criteria indicative of non-convergence can be tested for and ruled out, using for example a batch means test~\cite{2006-Jones-fixed_MCMC}. 
The MCMC procedure leads to a sample-based posterior distribution, from which the statistical behavior of the stochastic model can be easily inferred (for more details see Ref.~\citenum{mebane2013PCCP}.

\subsection{Exchange-Correlation Functionals}
XC functionals play a vital role in DFT. Numerous attempts have been made in the past to model the XC functional for accurate prediction of many-body quantum interactions~\cite{Lib_xc_Marques_2012, libxc_recent_LEHTOLA20181}. In particular, the precise description of the metal-to-insulator transition in strongly correlated materials requires methods that go further than a single determinant of the N-electron wave function~\cite{DFT+U_LDA+U}. Even though DFT is an exact theory, the perfect XC functional is not yet known.

The local density approximation (LDA), proposed by Kohn and Sham~\cite{kohn1965self}, adopts the exchange and correlation energies of the homogeneous electron gas~\cite{LDA,LDA-jones1989density,LDA_ceperley1980ground,LSDA_vonBarth}. It follows that LDA is most successful in predicting the properties of solids whose effects of exchange and correlation are short-range~\cite{LDA-jones1989density}. Nevertheless, it is broadly used in different material classes. LDA is known to underestimate exchange energy and overestimate correlation energy~\cite{gupta_principles_2015}. LDA systematically overbinds atoms causing an underestimation of the bond lengths and lattice parameters. 

Generalized-gradient approximation (GGA) XC are semi-local functionals that consider the gradient electron density to account for the anisotropic manner of the localized electron densities~\cite{PBE-PRB,PBE-PRL} of many materials. Contrary to LDA, GGA functionals tend to underbind atoms overestimating bond lengths and lattice constants. Perdew-Burke-Ernzerhof (PBE)~\cite{PBE-PRB,PBE-PRL} is the most popular GGA XC functional and has been used successfully to study many types of materials~\cite{wentzcovitch_theoretical_2018}. 

Similar to PBE, Perdew-Burke-Ernzerhof revised for solids (PBEsol)~\cite{PBEsol,PBEsol_PhysRevLett.100.136406} is a GGA XC functional. PBEsol differs from PBE only by two altered parameters that allow PBEsol to maintain many of the reliable properties from PBE~\cite{PBEsol_PhysRevLett.100.136406}. PBEsol improves the equilibrium properties such as bond lengths and lattice parameters over PBE. However, it is generally poor in predicting dissociation or cohesive energies and reaction energy barriers~\cite{PBE_PBEsol_structuralcomp_dongho_nguimdo_density_2015,zhang2018performance,de2011performance, hinuma2017comparison}. 

\subsection{DFT+$U$}

The correction in DFT for strongly correlated materials can be introduced by including the Hubbard model~\cite{liechenstein_PRB}.

\begin{equation}                            
    E_{DFT+U}[\rho^\sigma(r),\{n^{i\sigma}_{mm^\prime}\}]=E_{DFT}[\rho(r)]+E_{Hub}[\{n^{i\sigma}_{mm^\prime}\}] - E_{dc}[\{n^{i\sigma}_{mm^\prime}\}],
\end{equation}

where $\rho^\sigma(r)$ represents the charge density for spin $\sigma$ and $n^{i\sigma}_{mm^\prime}$ represents the density matrix for site $i$, states $m$ and $m^\prime$, and spin $\sigma$. The $E_{Hub}$ is the Hubbard correction for the electron-electron interaction that is only applied to specified correlated states ($d-$ and $f-$electrons). The $E_{dc}$, known as the double counting term, contains the energy of the correlated electrons calculated within DFT~\cite{ryee2018effect,wehling20145}. This term must be subtracted from the total energy as the Hubbard term already contains the corrected energy of these states. The $E_{Hub}$ used in this study is the rotationally invariant form introduced by Lichtenstein \textit{et al.}~\cite{liechenstein_PRB}. In this form, the Hubbard Hamiltonian is written in terms of matrix elements of the Coulomb electron-electron interaction. The matrix elements can be expanded in terms of Slater integrals and spherical harmonics. The effective Coulomb and exchange interactions, $U$ and $J$ are defined using the matrix elements of the Coulomb electron-electron interaction. Using atomic orbitals to extract the Slater integrals can lead to a large overestimation because the Coulomb interaction is screened. In DFT simulation packages, $U$ and $J$ are treated as parameters to reach an agreement with experimental results.

The DFT+$U$ method offers a relatively simple solution to the complex problem of XC interaction calculation in strongly correlated materials. In this work, the method used to determine the double-counting correction in the Hubbard Hamiltonian was the rotationally invariant method proposed by Liechtenstein~\cite{liechenstein_PRB}.

\subsection{Studied Materials} \label{sec:studied_materials}
 
In this study, we experimented with a group of iron-based compounds \ce{Fe} (Im$\overline{3}$m), \ce{Fe3Ge} (P6$_{3}$/mmc), \ce{Fe2P} (P$\overline{6}$2m), \ce{SrFeO3} (Pm$\overline{3}$m) and \ce{BaFeO3} (Pm$\overline{3}$m) having different Fe oxidation states. The experimental properties and crystal structures of each material are listed in Table~\ref{tab:experimental_data}. For Fe, \ce{BaFeO3}, and \ce{SrFeO3} we chose the cubic phases, while for \ce{Fe3Ge} and \ce{Fe2P}, we chose their hexagonal phase. In our calculations, Fe, \ce{BaFeO3}, \ce{SrFeO3}, \ce{Fe3Ge}, and \ce{Fe2P} have two, five, five, eight, and nine atoms per unit cell, respectively. Fe, \ce{Fe3Ge}, and \ce{Fe2P} have a ferromagnetic (FM) ordering~\cite{greenwood2012chemistry, Severin_1995, Drijver_1976}, while \ce{BaFeO3} and \ce{SrFeO3} exhibit a helimagnetic (HM) ordering~\cite{hayashi_BaFeO3}. 

\ce{SrFeO3} is a cubic perovskite and its HM structure propagates along $<$111$>$ direction by 46$^{\circ}$ from one layer to another~\cite{hayashi_BaFeO3}. Zhao and Zhou~\cite{zhao_SrFeO3} suggest that at low temperatures \ce{SrFeO3} adopts domains of FM phase causing magnetic inhomogeneity generating a metal-to-insulator transition. Given that our study is for 0 K, we use the FM ordered \ce{SrFeO3} phase.

As for \ce{BaFeO3}, it is well known that depending on the oxygen deficiency and temperature, it can adopt different crystal structures including triclinic, rhombohedral, tetragonal, and cubic~\cite{mori_BaFeO3,hayashi_BaFeO3}. These different phases correspond to different magnetic orderings ranging from the HM in the hexagonal to the FM in the cubic phase~\cite{hayashi_BaFeO3, nortonBaFeO3_report}. This material is reported to be an insulator in the cubic phase~\cite{BaFeO3_mag_band_PRB}. \ce{BaFeO3} follows the $<$100$>$ magnetic propagation direction and the helical structure rotates the y-z component of the spin by 22$^{\circ}$. Based on this smaller angle, \ce{BaFeO3} is closer to a ferromagnetic structure than \ce{SrFeO3}~\cite{hayashi_BaFeO3}. This is supported by the large magnetic field (42 T)~\cite{ishiwata_SrFeO3} required to switch \ce{SrFeO3} from HM to FM compared to the small magnetic field (0.3 T)~\cite{BaFeO3_mag_band_PRB} required to switch \ce{BaFeO3}. Given the small HM characteristic turn angle in the \ce{BaFeO3}, we considered this structure to be FM for this investigation.

We performed our calculations assuming that all structures had a collinear FM ordering. This assumption was made considering computational efficiency. Moreover, both of the perovskites were assumed to be insulating and in their cubic phases. Even though \ce{SrFeO3} is not insulating, we purposefully selected a bandgap for this material (we choose a bandgap reported for a thin film~\cite{gap_SrFeO3}, to both evaluate the robustness of MCMC to errors in small target values and avoid overfitting towards metallic states.

Using the MCMC sampling, the space of $U$ and $J$ parameters was built up with the calculations made for these five compounds. The mean values of the of $U$ and $J$ parameters were extracted from the estimated distribution after the burn-in. Using these mean values, we performed simulations for the original five materials as well as for the new materials: FeO (Fm$\bar{3}$m), \ce{$\alpha$-Fe2O3} (R$\bar{3}$c), \ce{Al2FeB2} (Cmmm), \ce{Fe5PB2} (I4/mcm), and \ce{Fe5SiB2} (I4/mcm).

\section{Results and discussion} \label{sec:results_and_discussion}

For each XC functional, we see that after a certain critical number of pairs of proposed parameters, equilibration (burn-in) is reached, and the algorithm starts to efficiently explore the most important regions of parameter space. The critical number of proposed parameters are approximately two-thousand pairs for PBE and PBEsol, and fifteen-hundred pairs for LDA. LDA and PBEsol explored different areas of parameter space more frequently than PBE. The progression of parameters is provided in MCMC trace plots in supplementary Figure 1.

The Hubbard model was introduced to DFT to correct the errors in the simplifications of the XC functionals. However, these corrections can be system dependent. Therefore, if the distribution of the correction parameters applied to various materials is localized, one can conclude that the correction parameters can be used universally in that specific XC functional with similar materials with reasonably good accuracy. 

After the PBE+$U$ Markov Chain reached the stationary zone (\textit{ca.} 2500 pairs of proposed $U$ and $J$), the parameters varied minimally until it was terminated (\textit{ca.} 8000 pairs). This leads us to believe that once the critical number of proposed pairs is reached and the algorithm locates an initial minimal variance of proposed parameters, it will not locate another in parameter space. The same behavior was observed for LDA+$U$ and PBEsol+$U$. This suggests that there is only one maximum for the $U$ and $J$ probability density distribution.

The arithmetic means and standard deviations of the $U$ and $J$ parameters are displayed in Table~\ref{tab:univariate_analysis}. The standard deviation of the $J$ parameter ($\sigma_J$) is smaller than that of $U$ ($\sigma_U$) for all three XC functionals. This is due to the higher effect the Coulomb-repulsion has on the energetics of a system compared to the exchange interaction. The mean value of the $J$ parameter ($J_{avg}$) is larger than the $J$ values used in other DFT+$U$ investigations \cite{csacsiouglu2011effective,ryee2018effect, bousquet2010j}. However, recent studies have shown that larger values of $J$ are needed to reproduce the magnetic moments of some iron compounds \cite{himmetoglu2014IJQC,nakamura2009first}. These larger values of $J$ tend to decrease the overprediction of the magnetic moment (See Supplementary Figure \ref{SI:fig:lda_heat_map}).

\begin{table}
\centering
\caption{Univariate analysis of the parameter space distributions. $U_{avg}$ and $J_{avg}$ represent the arithmetic mean of each distribution. $\sigma_U$ and $\sigma_J$ denote the standard deviation. $\sigma_{UJ}$ denotes the overall standard deviation. Lastly $\rho_{UJ}$ represents the Pearson correlation coefficient between $U$ and $J$ parameters.}
\vspace{0.5cm}
\begin{tabular}{l@{\hskip 0.6in}llll}
\hline
XC Functional &
  $U_\textrm{avg}$ ($\sigma_U$) &
  $J_\textrm{avg}$ ($\sigma_J$) &
  $\sigma_{UJ}$ &
  $\rho_{UJ}$ \\ \hline
LDA    & 5.9 (1.0) & 2.1 (0.6) & 1.4 & 0.5 \\
PBE    & 3.1 (0.3) & 1.9 (0.1) & 0.1 & 0.7 \\
PBEsol & 4.5 (0.6) & 2.1 (0.4) & 0.5 & 0.2 \\ \hline
\end{tabular}
\label{tab:univariate_analysis}
\end{table}

The mean value of the $U$ parameter ($U_{avg}$) is substantially larger in LDA in comparison to its GGA counterparts (PBE and PBEsol). This is expected as LDA is the simplest XC functional. As mentioned earlier, LDA assumes the XC energy is that of a homogenous electron gas. Therefore, it requires a greater on-site electron-electron Coulomb-interaction correction. LDA systematically overbinds the atoms causing an underestimation in the bond lengths. Thus, it requires a larger $U$ parameter to create the Coulomb-repulsion and expand the bonds and consequently the lattice parameters. Table~\ref{tab:experimental_data} shows this initial underestimation in the lattice parameters and the subsequent improvement when introducing $U$ and $J$ in the calculations. Regarding the GGA functionals, PBEsol required a slightly larger $U$ parameter than PBE. One of the purposes for the introduction of PBEsol was to correct the overestimation of PBE~\cite{PBEsol_PhysRevLett.100.136406} in the bond lengths for non-correlated materials. For correlated materials, however, this overestimation leads to a closer prediction in bond length to the experimentally measured because correlated materials need an extra Coulomb-repulsion for more precise predictions.

\begin{figure}[t]
    \centering
    \vspace*{-5cm}\includegraphics[width=16cm]{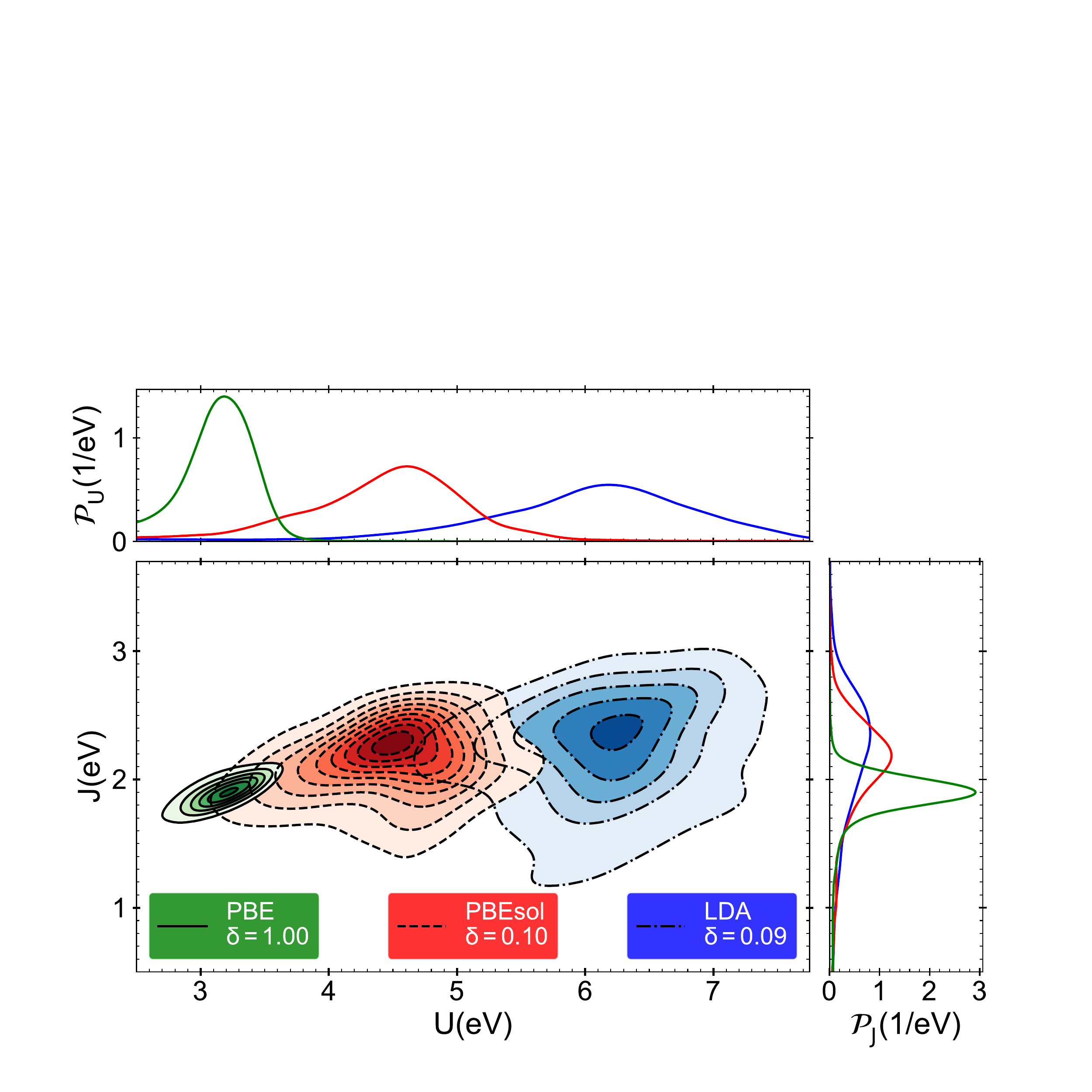}
    \caption{The density was estimated using a gaussian kernel density estimation (KDE). The bandwidth was selected using the Scott~\cite{KDE_scott} approach. Each KDE is normalized to one separately. $\delta$ is the step between contour lines. (top left) Shows the probability density function of accepted $U$ parameters. (bottom left) Shows the joint probability density function of accepted $U$ and $J$. (bottom right) Shows the probability density function of accepted $J$ parameters.}
    \label{fig:probDist}
\end{figure}

The distribution of $U$ and $J$ parameters is more localized in PBE comparing to that of LDA and PBEsol. This can be visualized in Figure~\ref{fig:probDist} by noting the spread of the distribution in the parameter space in each case. Furthermore, the univariate analysis, provided in Table~\ref{tab:univariate_analysis}, shows that PBE has a noticeably smaller overall standard deviation ($\sigma_{UJ}$) than LDA and PBEsol. A small overall standard deviation of $U$ and $J$ in the parameter space (\textit{i.e.} a localized distribution) indicates that using the mean values $U_{avg}$ and $J_{avg}$ simultaneously improves the results toward a better agreement with the experimental data for all of the structures. Therefore, we expect $U_{avg}$ and $J_{avg}$ values from the distribution for PBE+$U$ are more transferable to other materials than LDA+$U$ and PBEsol+$U$.

The last column of Table \ref{tab:univariate_analysis} shows the Pearson correlation coefficient of the $U$ and $J$ parameter ($\rho_{UJ}$). If the correlation factor is equal to zero, $U$ and $J$ are completely independent. As the correlation approaches one, the dependence increases. If the correlation is equal to one, $U$ and $J$ are completely dependent. This is reminiscent of the Dudarev approximation ~\cite{dudarev}, a more simplified yet rotationally invariant form, where the functional can be obtained by only considering the zeroth-order Slater integral. The treatment of $U$ and $J$ values in Ref. ~\citenum{dudarev} is analogous to incorporating the exchange interaction to the Coulomb interaction using an effective $U$, $U_{\mathrm{eff}} = U–J$ ~\cite{himmetoglu2014IJQC}. Within the Dudarev approximation the two parameters of Lichtenstein form, $U$ and $J$, are effectively reduced to one parameter, $U_{\mathrm{eff}}$. We find that PBE has the largest correlation between $U$ and $J$. This seems to indicate that out of the three studied XC functionals, PBE has the closest result between Dudarev approximation ~\cite{dudarev} and Lichtenstein form~\cite{liechenstein_PRB}.

We have recorded the experimental and predicted values of lattice parameters, volume, bandgap, and magnetic moment for the studied materials in Table~\ref{tab:experimental_data}. Even though volume, bandgap, and magnetic moment were set equally as target parameters, it can be seen that the corrections for lattice parameters have been more effective than the bandgap and magnetic moment. This is because treating the volume on the same footing as bandgap and magnetic moment increases the importance of the lattice parameters. Also, changes in lattice parameters can subsequently effect the magnetic moment and bandgap predictions.

\begin{table}[]
\caption{Structural, electronic, and magnetic properties of selected iron-based compounds. Values outside \textcolor{blue}{(inside)} parenthesis are from simulations using DFT \textcolor{blue}{(DFT+$U$)}. The DFT+$U$ calculations were performed using the mean values of $U$ and $J$ from the distributions. Letters a, b, and c represent the lattice parameters. MP represents the final magnetic phase. 
Volume, bandgap, and magnetic moment are expressed in units of \r{A}$^3$, eV, and Bohr magneton ($\mu_B$), respectively.}
\label{tab:experimental_data}
\resizebox{\textwidth}{!}{
\begin{tabular}{lllllllll}
\hline \hline
Material &
  XC &
  a &
  b &
  c &
  Volume &
  Bandgap &
  Mag. Mom. &
  MP \\ \hline \hline
\ce{Fe} &
  Experiment &
  2.87\textsuperscript{\emph{a}} &
   &
   &
  23.64 &
  0.0\textsuperscript{\emph{b}} &
  2.22\textsuperscript{\emph{c}} &
  FM\textsuperscript{\emph{b}} \\
Im$\overline{3}$m &
  LDA \textcolor{blue}{(+$U$)} &
  2.75 \textcolor{blue}{(2.83)} &
   &
   &
  20.71 \textcolor{blue}{(22.55)} &
  0.00 \textcolor{blue}{(0.00)} &
  1.95 \textcolor{blue}{(2.73)} &
  FM \textcolor{blue}{(FM)} \\
 &
  PBE, \textcolor{blue}{(+$U$)} &
  2.83 \textcolor{blue}{(2.84)} &
   &
   &
  22.58 (\textcolor{blue}{22.96}) &
  0.00 (\textcolor{blue}{0.00}) &
  2.19 (\textcolor{blue}{2.09}) &
  FM \textcolor{blue}{(FM)} \\
 &
  PBEsol \textcolor{blue}{(+$U$)} &
  2.78 \textcolor{blue}{(2.85)} &
   &
   &
  21.59 \textcolor{blue}{(23.22)} &
  0.00 \textcolor{blue}{(0.00)} &
  2.12 \textcolor{blue}{(2.71)} &
  FM \textcolor{blue}{(FM)} \\ \hline
\ce{Fe2P} &
  Experiment &
  5.87\textsuperscript{\emph{d}} &
   &
  3.46\textsuperscript{\emph{d}} &
  119.34 &
  0.0\textsuperscript{\emph{e}} &
  1.91\textsuperscript{\emph{f}} (Fe(II)) &
  FM\textsuperscript{\emph{f}} \\
P$\overline{6}$2m &
  LDA \textcolor{blue}{+$U$)} &
  5.56 \textcolor{blue}{(5.88)} &
   &
  3.42 \textcolor{blue}{(3.32)} &
  91.31 \textcolor{blue}{(99.37)} &
  0.00 \textcolor{blue}{(0.00)} &
  1.11 \textcolor{blue}{(2.26)} &
  FM \textcolor{blue}{(FM)} \\
 &
  PBE \textcolor{blue}{(+$U$)} &
  5.81 \textcolor{blue}{(5.91)} &
   &
  3.41 \textcolor{blue}{(3.38)} &
  99.55 \textcolor{blue}{(102.43)} &
  0.00 \textcolor{blue}{(0.00)} &
  2.25 \textcolor{blue}{(2.09)} &
  FM \textcolor{blue}{(FM)} \\
 &
  PBEsol \textcolor{blue}{(+$U$)} &
  5.70 \textcolor{blue}{(5.90)} &
   &
  3.40 \textcolor{blue}{(3.36)} &
  95.70 \textcolor{blue}{(101.25)} &
  0.00 \textcolor{blue}{(0.00)} &
  2.03 \textcolor{blue}{(2.23)} &
  FM \textcolor{blue}{(FM)} \\ \hline
\ce{Fe3Ge} &
  Experiment &
  5.17\textsuperscript{\emph{g}} &
   &
  4.22\textsuperscript{\emph{g}} &
  112.79 &
  0.0\textsuperscript{\emph{g}} &
  2.00\textsuperscript{\emph{g}} &
  FM\textsuperscript{\emph{g}} \\
P6$_{3}$/mmc &
  LDA \textcolor{blue}{(+$U$)} &
  4.95 \textcolor{blue}{(5.18)} &
   &
  4.03 \textcolor{blue}{(4.17)} &
  85.69 \textcolor{blue}{(96.80)} &
  0.00 \textcolor{blue}{(0.00)} &
  1.25 \textcolor{blue}{(2.75)} &
  FM \textcolor{blue}{(FM)} \\
 &
  PBE \textcolor{blue}{(+$U$)} &
  5.14 \textcolor{blue}{(5.17)} &
   &
  4.20 \textcolor{blue}{(4.21)} &
  95.83 \textcolor{blue}{(97.50)} &
  0.00 \textcolor{blue}{(0.00)} &
  2.18 \textcolor{blue}{(2.37)} &
  FM \textcolor{blue}{(FM)} \\
 &
  PBEsol \textcolor{blue}{(+$U$)} &
  5.15 \textcolor{blue}{(5.17)} &
   &
  4.22 \textcolor{blue}{(4.28)} &
  96.85 \textcolor{blue}{(98.97)} &
  0.00 \textcolor{blue}{(0.00)} &
  2.17 \textcolor{blue}{(2.66)} &
  FM \textcolor{blue}{(FM)} \\ \hline
\ce{BaFeO3} &
  Experiment &
  3.97\textsuperscript{\emph{h}} &
   &
   &
  62.57 &
  1.8\textsuperscript{\emph{i}} &
  3.50\textsuperscript{\emph{i}} &
  FM\textsuperscript{\emph{i}} \\
Pm$\overline{3}$m &
  LDA \textcolor{blue}{(+$U$)} &
  3.86 \textcolor{blue}{(3.90)} &
   &
   &
  57.31 \textcolor{blue}{(59.09)} &
  0.00 \textcolor{blue}{(0.00)} &
  2.64 \textcolor{blue}{(3.56)} &
  FM \textcolor{blue}{(FM)} \\
 &
  PBE \textcolor{blue}{(+$U$)} &
  3.97 \textcolor{blue}{(3.98)} &
   &
   &
  62.47 \textcolor{blue}{(63.24)} &
  0.00 \textcolor{blue}{(0.00)} &
  3.02 \textcolor{blue}{(3.37)} &
  FM \textcolor{blue}{(FM)} \\
 &
  PBEsol \textcolor{blue}{(+$U$)} &
  3.90 \textcolor{blue}{(3.91)} &
   &
   &
  59.39 \textcolor{blue}{(59.75)} &
  0.00 \textcolor{blue}{(0.00)} &
  2.88 \textcolor{blue}{(3.45)} &
  FM \textcolor{blue}{(FM)} \\ \hline
\ce{SrFeO3} &
  Experiment &
  3.85\textsuperscript{\emph{j}} &
   &
   &
  57.06 &
  1.8\textsuperscript{\emph{k}} &
  3.10\textsuperscript{\emph{m}} &
  FM\textsuperscript{\emph{o}} \\
Pm$\overline{3}$m &
  LDA \textcolor{blue}{(+$U$)} &
  3.74 \textcolor{blue}{(3.78)} &
   &
   &
  52.24 \textcolor{blue}{(53.93)} &
  0.00 \textcolor{blue}{(0.00)} &
  2.51 \textcolor{blue}{(3.49)} &
  FM \textcolor{blue}{(FM)} \\
 &
  PBE \textcolor{blue}{(+$U$)} &
  3.84 \textcolor{blue}{(3.85)} &
   &
   &
  56.70 \textcolor{blue}{(57.21)} &
  0.00 \textcolor{blue}{(0.00)} &
  2.87 \textcolor{blue}{(3.15)} &
  FM \textcolor{blue}{(FM)} \\
 &
  PBEsol \textcolor{blue}{(+$U$)} &
  3.77 \textcolor{blue}{(3.79)} &
   &
   &
  53.45 \textcolor{blue}{(54.64)} &
  0.00 \textcolor{blue}{(0.00)} &
  2.71 \textcolor{blue}{(3.36)} &
  FM \textcolor{blue}{(FM)} \\ \hline \hline
\ce{FeO} &
  Experiment &
  4.31\textsuperscript{\emph{q,s}} &
   &
   &
  80.06 &
  1\textsuperscript{\emph{p}}-2.4\textsuperscript{\emph{r}} &
  3.32\textsuperscript{\emph{q}} &
  AFM\textsuperscript{\emph{q}} \\
Fm$\overline{3}$m &
  LDA \textcolor{blue}{(+$U$)} &
  4.15 \textcolor{blue}{(4.20)} &
   &
   &
  71.28 \textcolor{blue}{(73.31)} &
  0.00 \textcolor{blue}{(2.85)} &
  3.30 \textcolor{blue}{(0.12)} &
  AFM \textcolor{blue}{(AFM)} \\
 &
  PBE \textcolor{blue}{(+$U$)} &
  4.24 \textcolor{blue}{(4.27)} &
   &
   &
  76.43 \textcolor{blue}{(77.74)} &
  0.00 \textcolor{blue}{(0.00)} &
  3.40 \textcolor{blue}{(3.51)} &
  AFM \textcolor{blue}{(AFM)} \\
 &
  PBEsol \textcolor{blue}{(+$U$)} &
  4.15 \textcolor{blue}{(4.22)} &
   &
   &
  70.25 \textcolor{blue}{(75.24)} &
  0.00 \textcolor{blue}{(0.00)} &
  3.29 \textcolor{blue}{(3.55)} &
  AFM \textcolor{blue}{(AFM)} \\ \hline
\ce{$\alpha-$Fe2O3} &
  Experiment &
  5.03\textsuperscript{\emph{t}} &
   &
  13.75\textsuperscript{\emph{t}} &
  301.82 &
  2.1\textsuperscript{\emph{u}} &
  4.9\textsuperscript{\emph{u}} &
  AFM\textsuperscript{\emph{v}} \\
R$\overline{3}$c &
  LDA \textcolor{blue}{(+$U$)} &
  4.62 \textcolor{blue}{(4.95)} &
   &
  13.31 \textcolor{blue}{(13.60)} &
  246.03 \textcolor{blue}{(289.03)} &
  0.00 \textcolor{blue}{(1.74)} &
  1.11 \textcolor{blue}{(4.00)} &
  AFM \textcolor{blue}{(AFM)} \\
 &
  PBE, \textcolor{blue}{(+$U$)} &
  5.00 \textcolor{blue}{(5.05)} &
   &
  13.86 \textcolor{blue}{(13.91)} &
  300.59 \textcolor{blue}{(306.85)} &
  0.53 \textcolor{blue}{(1.15)} &
  3.55 \textcolor{blue}{(3.85)} &
  AFM \textcolor{blue}{(AFM)} \\
 &
  PBEsol \textcolor{blue}{(+$U$)} &
  4.91 \textcolor{blue}{(5.00)} &
   &
  13.66 \textcolor{blue}{(13.73)} &
  285.18 \textcolor{blue}{(297.22)} &
  0.30 \textcolor{blue}{(1.49)} &
  3.36 \textcolor{blue}{(3.95)} &
  AFM \textcolor{blue}{(AFM)} \\ \hline
\ce{AlFeB2} &
  Experiment &
  2.92\textsuperscript{\emph{w}} &
  11.03\textsuperscript{\emph{w}} &
  2.87\textsuperscript{\emph{w}} &
  92.23\textsuperscript{\emph{w}} &
  0.0\textsuperscript{\emph{x}} &
  1.21\textsuperscript{\emph{w,y,z}} &
  FM\textsuperscript{\emph{w,y,z}} \\
Cmmm &
  LDA \textcolor{blue}{(+$U$)} &
  2.90 \textcolor{blue}{(2.87)} &
  11.13 \textcolor{blue}{(10.84)} &
  2.64 \textcolor{blue}{(2.85)} &
  85.17 \textcolor{blue}{(88.53)} &
  0.0 \textcolor{blue}{(0.0)} &
  0.0 \textcolor{blue}{(1.64)} &
  FM \textcolor{blue}{(FM)} \\
 &
  PBE \textcolor{blue}{(+$U$)} &
  2.92 \textcolor{blue}{(2.92)} &
  11.01 \textcolor{blue}{(11.01)} &
  2.86 \textcolor{blue}{(2.86)} &
  91.91 \textcolor{blue}{(91.91)} &
  0.0 \textcolor{blue}{(0.0)} &
  1.40 \textcolor{blue}{(1.52)} &
  FM \textcolor{blue}{(FM)} \\
 &
  PBEsol \textcolor{blue}{(+$U$)} &
  2.92 \textcolor{blue}{(2.92)} &
  11.01 \textcolor{blue}{(11.01)} &
  2.86 \textcolor{blue}{(2.86)} &
  91.91 \textcolor{blue}{(91.91)} &
  0.0 \textcolor{blue}{(0.0)} &
  1.37 \textcolor{blue}{(1.57)} &
  FM \textcolor{blue}{(FM)} \\ \hline
\ce{Fe5PB2} &
  Experiment &
  5.49\textsuperscript{\emph{l}} &
   &
  10.35\textsuperscript{\emph{l}} &
  311.67 &
  0.0\textsuperscript{\emph{}} &
  1.73 \textsuperscript{\emph{l}} &
  FM\textsuperscript{\emph{l}} \\
I4/mcm &
  LDA \textcolor{blue}{(+$U$)} &
  5.45 \textcolor{blue}{(5.45)} &
   &
  10.31 \textcolor{blue}{(10.31)} &
  306.45 \textcolor{blue}{(306.45)} &
  0.0 \textcolor{blue}{(0.0)} &
  1.43 \textcolor{blue}{(2.21)} &
  FM \textcolor{blue}{(FM)} \\
 &
  PBE \textcolor{blue}{(+$U$)} &
  5.44 \textcolor{blue}{(5.51)} &
   &
  10.34 \textcolor{blue}{(10.39)} &
  305.79 \textcolor{blue}{(315.32)} &
  0.0 \textcolor{blue}{(0.0)} &
  1.79 \textcolor{blue}{(1.99)} &
  FM \textcolor{blue}{(FM)} \\
 &
  PBEsol \textcolor{blue}{(+$U$)} &
  5.35 \textcolor{blue}{(5.48)} &
   &
  10.18 \textcolor{blue}{(10.26)} &
  292.08 \textcolor{blue}{(308.12)} &
  0.0 \textcolor{blue}{(0.0)} &
  1.55 \textcolor{blue}{(2.11)} &
  FM \textcolor{blue}{(FM)} \\ \hline
\ce{Fe5SiB2} &
  Experiment &
  5.55\textsuperscript{\emph{l}} &
   &
  10.34\textsuperscript{\emph{l}} &
  318.45 &
  0.0\textsuperscript{\emph{}} &
  1.83\textsuperscript{\emph{l}} &
  FM\textsuperscript{\emph{l}} \\
I4/mcm &
  LDA \textcolor{blue}{(+$U$)} &
  5.45 \textcolor{blue}{(5.45)} &
   &
  10.31 \textcolor{blue}{(10.31)} &
  306.45 \textcolor{blue}{(306.45)} &
  0.00 \textcolor{blue}{(0.0)} &
  1.48 \textcolor{blue}{(2.11)} &
  FM \textcolor{blue}{(FM)} \\
 &
  PBE \textcolor{blue}{(+$U$)} &
  5.50 \textcolor{blue}{(5.54)} &
   &
  10.33 \textcolor{blue}{(10.42)} &
  312.25 \textcolor{blue}{(320.29)} &
  0.00 \textcolor{blue}{(0.0)} &
  1.84 \textcolor{blue}{(1.98)} &
  FM \textcolor{blue}{(FM)} \\
 &
  PBEsol \textcolor{blue}{(+$U$)} &
  5.43 \textcolor{blue}{(5.51)} &
   &
  10.12 \textcolor{blue}{(10.27)} &
  298.58 \textcolor{blue}{(312.42)} &
  0.0 \textcolor{blue}{(0.0)} &
  1.61 \textcolor{blue}{(2.04)} &
  FM \textcolor{blue}{(FM)} \\ 
 \hline \hline
\end{tabular}}
\textsuperscript{\emph{a}} Ref.~\citenum{Fe_structure_LandoltBornstein1994}; 
\textsuperscript{\emph{b}} Ref.~\citenum{cornell2004FeOxides_ch6}; 
\textsuperscript{\emph{c}} Ref.~\citenum{Fe_magMom};
\textsuperscript{\emph{d}} Ref.~\citenum{TOBOLA1996708};
\textsuperscript{\emph{e}} Ref.~\citenum{gap_Fe2P_SUGIZAKI201750};
\textsuperscript{\emph{f}} Ref.~\citenum{Severin_1995};
\textsuperscript{\emph{g}} Ref.~\citenum{Drijver_1976};
\textsuperscript{\emph{h}} Ref.~\citenum{BaFeO3_lattice_Taib2016StructuralEA};
\textsuperscript{\emph{i}} Ref.~\citenum{BaFeO3_mag_band_PRB};
\textsuperscript{\emph{j}} Ref.~\citenum{lattice_SrFeO3_osti_1376467};
\textsuperscript{\emph{k}} Ref.~\citenum{gap_SrFeO3};
\textsuperscript{\emph{l}} Ref.~\citenum{Fe5PB2mcguireFe5SiB2};
\textsuperscript{\emph{m}} Ref.~\citenum{MATAR1995169};
\textsuperscript{\emph{n}} Ref.~\citenum{lu_room_2010};
\textsuperscript{\emph{o}} Ref.~\citenum{2016ApPhA_SrFeO3};
\textsuperscript{\emph{p}} Ref.~\citenum{schrettle2012wustite};
\textsuperscript{\emph{q}} Ref.~\citenum{goodenough_magnetic_1970};
\textsuperscript{\emph{r}} Ref.~\citenum{bowen1975wustite};
\textsuperscript{\emph{s}} Ref.~\citenum{cornell2004FeOxides_ch2};
\textsuperscript{\emph{t}} Ref.~\citenum{1980-Finger-Fe2O3};
\textsuperscript{\emph{u}} Ref.~\citenum{Coey_1971};
\textsuperscript{\emph{v}} Ref.~\citenum{coey2013_mag_oxide};
\textsuperscript{\emph{w}} Ref.~\citenum{lamichhane2018AlFe2B2_mag_lat};
\textsuperscript{\emph{x}} Ref.~\citenum{barua2019AlFe2B2_conductor};
\textsuperscript{\emph{y}} Ref.~\citenum{elmassalami2011AlFe2B2_mag};
\textsuperscript{\emph{z}} Ref.~\citenum{ali2017AlFe2B2_mag}.
\end{table}

We selected an accuracy criterion of 0.09 \r{A} and compared the experimental and predicted lattice parameters before and after the Hubbard correction. As expected, LDA usually underestimates the lattice parameters. This corroborates our previous findings that LDA needs a larger $U$ value to correct the underestimation of the bond lengths. The introduction of the correction parameters improves the prediction for most of the structures. As mentioned before, PBE is known for overestimating lattice parameters in non-correlated materials. For strongly correlated materials, as in the case of this study, this trend benefits PBE in predicting the lattice parameters reasonably accurately without any corrections. This was also observed by Meng \textit{et  al.}~\cite{meng2016Fe_oxides} in their study of a group of iron oxides using beyond-DFT approaches, where they observed adding the suggested $U$ and $J$ parameters to PBE minimally influence the lattice parameters prediction. This result also supports our previous observation that PBE requires smaller correction parameters. On the other hand, PBEsol underestimates the lattice parameters. This was expected because PBEsol was introduced to correct the overestimation of PBE. The $U$ and $J$ parameters suggested in this study improve the lattice parameter prediction in PBEsol. Detailed analysis can be found in supplementary Table 2.

The same analysis was performed for the magnetic moment with an accuracy criterion of 0.2 $\mu_B$. Magnetic moment predictions by LDA are underestimated for all of the structures. This underestimation frequently turns to an overestimation by introducing the correctional parameters. PBE, however, usually predicts the magnetic moment accurately, and adding the suggested $U$ and $J$ does not change the number of accurate predictions. PBEsol, similar to LDA, underestimates the magnetic moment. The suggested correctional parameters convert this underestimation to overestimation. Detailed analysis can be found in supplementary Table 3.

As for bandgap predictions, predicting a zero bandgap by DFT+$U$ is not remarkable. The materials listed with a bandgap in Table~\ref{tab:experimental_data} are \ce{BaFeO3}, \ce{SrFeO3}, \ce{FeO}, and \ce{$\alpha$-Fe2O3}.\ce{BaFeO3} exhibits a metallic behavior even after the Hubbard correction. Additional calculations were performed with the aim to open the bandgap in this compound using higher values of $U$. However, this was not achieved, even with values as high as 8 eV. Similarly, \ce{SrFeO3} also shows a metallic behavior with and without the correctional parameters. Experimentally it has both metallic and insulating phases~\cite{zhao_SrFeO3}. To be able to capture the insulating phase using DFT one has to prepare a structure that includes both HM and FM domains. For \ce{FeO} (w{\"u}stite), the only XC functional that could open a bandgap using the Hubbard correction was LDA, however, the magnetic moment was drastically underestimated. 
Prediction of the correct bandgap in \ce{FeO} requires special care associated with the occupancies of the 3$d$ states~\cite{Cococcioni_PhysRevB.71.035105}.
Mandal \textit{et al.}~\cite{FeO_Subhasish, FeO_Subhasish_nature} showed DFT+$U$ is not sufficient for reproducing the experimental results of \ce{FeO} and one has to employ DFT+DMFT~\cite{Vollhardt_2011,DFTwDMFT_uthpala,dft_dmft1,kent2018dft_dmft, haule2015dft_dmft1, haule2015dft_dmft2, koccer2020dft_dmft, aichhorn2016dft_dmft, vollhardt2019dft_dmft} method to accurately predict the AFM state of \ce{FeO}. As for \ce{$\alpha$-Fe2O3} (hematite), before introducing $U$ and $J$ parameters, LDA predicted a metallic behavior, while PBE and PBEsol opened a small bandgap. Using the correctional parameters all three XC functionals estimated an acceptable bandgap without compromising other properties.

Finally, we show the root mean square error (RMSE) and mean absolute error (MAE) of the predicted properties (volume, magnetic moment) in Supplementary Table \ref{SI:tab:rmse_mae}. The RMSE and MAE show the improvement in the predicted values in all of the XC functionals after including the Hubbard correction. 

In summary, we selected a group of iron-based compounds and explored the space of the correction parameters $U$ and $J$ that can improve the prediction results (volume, magnetic moment, and bandgap) for all of the studied materials simultaneously. This semi-empirical exploration was done using a Bayesian calibration, assisted by Markov Chain Monte Carlo sampling. For these iron-based compounds, we extracted three sets of $U$ and $J$ for LDA, PBE, and PBEsol XC functionals. 
All the $U$ and $J$ distributions have a single maximum. 
LDA requires a significantly larger $U$ parameter comparing to GGA functionals. 
$U$ and $J$ achieved in PBE are the most transferable between the studied iron-based compounds. The Dudarev approximation can result in a closer prediction to the Lichtenstein form of the Hubbard interaction in PBE compared to that of LDA and PBEsol. Assessing the correction parameters obtained from the distributions, showed the suggested correctional parameters improve the prediction of the lattice parameters and the magnetic moment in all XC functionals. A correct bandgap was not predicted for \ce{FeO} or \ce{BaFeO3}, due to the inability of DFT+$U$ to reproduce the experimental results. In the case of \ce{$\alpha-$Fe2O3}, bandgap estimation was improved for all the XC functionals. PBE predicts the lattice parameters reasonably accurately even without the Hubbard correction for these iron-based compounds. Lastly, based on the analysis performed in this study, we conclude that the $U$ and $J$ pairs provided can be a good starting point for DFT+$U$ calculations on the iron-based compound. In the  future, it will be interesting to expand the parameter space to incorporate the details of the orbital occupation~\cite{MeredigPRB2010, AllenWatson2014,payne2019PCCP}, the inter-site Hubbard $V$~\cite{himmetoglu2014IJQC}, and pseudopotentials~\cite{kulik-marzari2008pseudo}. Moreover, various other properties such as cohesive energy, formation energy, elastic constants, \textit{etc.} can be used in the dataset $X$. The proposed methodology can be employed for other systems to predict their properties for a given set of parameters within the spirit of high-throughput calculations.

\section{Methods} \label{sec:computational_details}
\subsection{DFT+$U$ and Bayesian calibration interface}
Since the underlying model is nonlinear and the evidence $P(X)$ is intractable, we used MCMC to draw the samples from the distribution. The MCMC sampler used an adaptive block proposal. For each run of the sampler, post equilibration (burn-in) convergence was assessed using a standard of $\pm$ 5\% for both $U$ and $J$ at 95\% confidence using a Student \textit{t}-test on batch means. Mixing of the sampler depicts a stationary behavior, and convergence was obtained for all runs after approximately 2000 post-burn-in draws.

\begin{figure}[h]
    \centering
    \includegraphics[width=10cm]{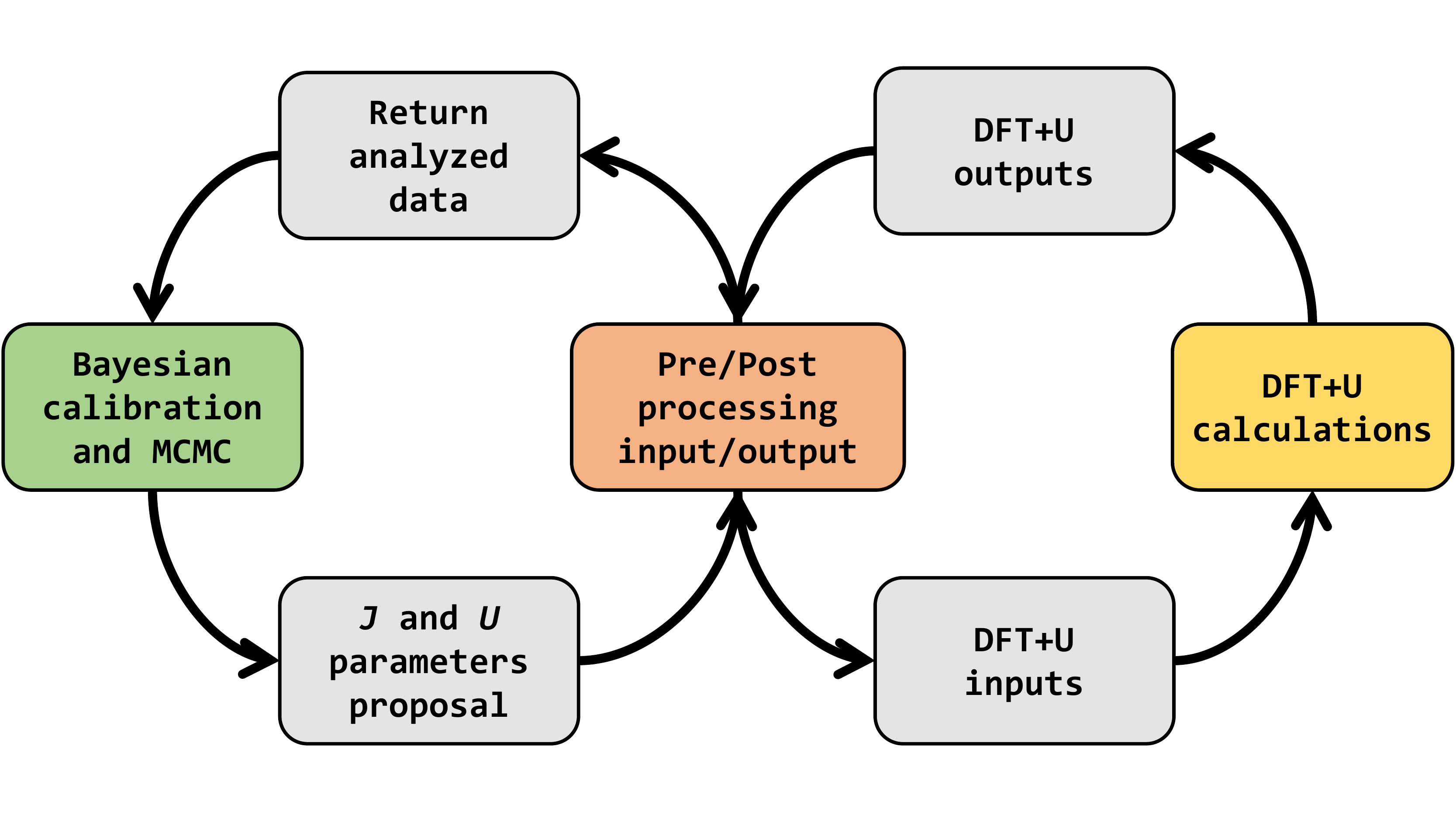}
    \caption{Computational process.}
    \label{fig:comp_process}
\end{figure}

This experiment is a set of back and forth communications between the DFT package and the MCMC sampler. The DFT+$U$ calculation is performed using the $U$ and $J$ parameters proposed by the MCMC sampler. Based on the accuracy of the DFT prediction in comparison with the experimental values, the MCMC sampler proposes a new pair of parameters drawn from a normal distribution centered at the $U$ and $J$ of the previous step for a new trial, and so on. We  use a block-proposal scheme (both \textit{i.e.} $U$ and $J$ are proposed at once). Our implementation uses an adaptive proposal where the covariance of the multivariate normal proposal distribution is shaped to the accepted points. At each MCMC step the likelihood is calculated and the proposal is accepted or rejected based on the Metropolis-Hastings algorithm (for more details see Ref. ~\citenum{mebane2013PCCP}. A schematic representation of the algorithm is shown in Figure~\ref{fig:comp_process}.

\subsection{Computational details}
The DFT calculations were performed using the Vienna Ab initio Simulation Package (VASP)~\cite{vasp1,vasp2,vasp3,vasp4}. The valence electrons wave functions were described by the projector augmented-wave~\cite{blochl_PAW,kresse_PAW} method. The kinetic energy expansion and optimum irreducible Brillouin zone grid (\textit{k}-grid) for each structure were obtained by choosing a maximum error of 1 meV/atom for the total energy in each cell. We used  $\Gamma$-centered and 
Monkhorst-Pack type~\cite{monkhorst} \textit{k}-grids
for hexagonal and cubic structures, respectively. Detailed convergence parameters are provided in supplementary Table 1. The Slater integrals values for Fe 3d shell were evaluated using the $U$, $J$, and the ratio of $F^4/F^2$, as implemented in VASP~\cite{bengone2000implementation}.

The Kohn-Sham equations were solved self-consistently with a maximum total energy difference of $10^{-5}$ eV. Furthermore, we assumed the crystal structure geometry to be optimized when the internal stress tensor components differ from the ambient pressure (assumed to be zero) by less than 0.5 kbar, and the residual forces on each atom are less than 1 meV/\r{A}.

As a consequence of the MCMC random walk, the algorithm might step in unphysical areas of the parameter space where $J > U$. These values are expected to be proposed because the Markov chain is free to explore every possible region seeking points where the predictions are close to the provided experimental values. Initially, the algorithm has little guidance from past proposed parameters leading to the proposition of unphysical parameters. To penalize the MCMC walker anytime an unphysical pair is proposed by the sampler, we skip the DFT calculation and return senseless values for the DFT+$U$ prediction (\textit{e.g.} bandgap = -50 eV, volume = -50 \r{A}$^3$, magnetic moment = -50 $\mu_B$). This encourages the algorithm to avoid proposing unnatural parameters and to explore other areas of the parameter space. The same strategy is used to penalize the algorithm when the $U$ and $J$ correction results in a change of space group.

\section*{Acknowledgements} \label{sec:acknowledgements}

This work used the XSEDE which is supported by the National Science Foundation (NSF) (ACI-1053575). The authors also acknowledge the support from the Texas Advanced Computing Center and the Pittsburgh Supercomputing Center (with the Stampede2 and Bridges supercomputers). We also acknowledge the use of the Thorny Flat Cluster at WVU, which is funded in part by the NSF Major Research Instrumentation Program (MRI) Award (MRI-1726534). Additionally, we acknowledge the support of O'Brien Fund of the WVU Energy Institute and the Summer Undergraduate Research Experience (SURE) at WVU. This work was supported by the DMREF-NSF 1434897, NSF OAC-1740111, and DOE DE-SC0021375 projects. Figures in this paper were generated using the Matplotlib~\cite{matplotlib} and PyVista~\cite{pyvista} python packages. We used Numpy~\cite{numpy} and SciPy~\cite{scipy} Python packages for pre- and post-processing of the results.

\section*{Supplementary information}

\setcounter{figure}{0}    
\setcounter{table}{0}    

\renewcommand{\figurename}{Supplementary Figure}
\renewcommand{\tablename}{Supplementary Table}

Supplementary Table \mbox{~\ref{SI:tab:rmse_mae}} shows the improvement in volume and magnetic moment prediction using DFT+$U$.

In Supplementary Table \mbox{~\ref{SI:tab:BaFe2As2}} for \ce{BaFe2As2} we used the \textit{Fmmm} space group and started the computation with ferromagnetic (FM), stripe antiferromagnetic (s-AFM), and checkerboard antiferromagnetic (c-AFM) initial magnetic phases. The DFT+$U$ calculations were performed using the mean values of $U$ and $J$ from the distributions. This material was selected to evaluate the performance of these suggested $U$ and $J$ values in an iron-based superconductor. It is known that this compound is an antiferromagnet below approximately 140 K. LDA, regardless of XC functional, predicts a nonmagnetic (NM) phase. The total energies using LDA for the different magnetic phases are very similar to each other and do not allow an accurate prediction of the ground state. LDA+$U$ predicts the ground state of this compound to be s-AFM with a magnetic moment of 2.94 $\mu_B$. PBE predicts the s-AFM phase to be the most stable magnetic phase with a magnetic moment of 1.43 $\mu_B$, and PBE+$U$ predicts the FM phase as the ground state with an overestimated magnetic moment of 2.17 $\mu_B$. This example clearly demonstrates that certain materials require a larger $J$ value in order to accurately describe their magnetic behavior. For this specific compound, it appears preferable not to use the Hubbard correction than to use a small $J$ value. Lastly, PBEsol predicts the ground state to be s-AFM for both PBEsol and PBEsol+$U$. Similar to other XC functionals, the Hubbard correction tends to overestimate the magnetic moment (2.86 $\mu_B$) while the plain calculation better approaches the experimental values. As shown by Derondeau \textit{et al.}, calculations for this material often overestimate magnetic moments\cite{derondeau2016hyperfine}.

Supplementary Figure \mbox{~\ref{SI:fig:lda_heat_map}} shows percentage error heat maps of $U$ and $J$ for \mbox{\ce{Fe}} and \mbox{\ce{SrFeO3}} for volume, bandgap, and magnetic moment predictions. The $U$ and $J$ pairs that provide the most accurate results are different for iron in different environments. The same can be said for the prediction of different properties in the same material, \textit{i.e.} the $U$ and $J$ pairs that produce the closes result for volume are not the same for magnetic moment. For simplicity, here we show all prediction errors only for \mbox{\ce{Fe}} and \mbox{\ce{SrFeO3}} in LDA+$U$. As plotted, we only show the pixels that were explored by the sampler. The $U$ and $J$ extracted from this sampling are shown as solid round magenta dots.

To evaluate the importance of target properties and the functionality of the method, we performed the experiment by omitting the bandgap from the target properties for LDA and PBE. This pushes the algorithm towards providing $U$ and $J$ pairs that improve the magnetic moment and the volume more than when bandgap is included in the target properties. The $U$ and $J$ parameters extracted from this sampling are provided in Supplementary Table \mbox{~\ref{SI:tab:noBG}}. The probability distribution for sampling is shown in Supplementary Figure \mbox{~\ref{SI:fig:distribution_noBG}}. Supplementary Table \mbox{~\ref{SI:tab:noBG_rmse}} shows the root mean square error (RMSE) and mean absolute error (MAE) from the DFT+$U$ calculation using the $U$ and $J$ pair extracted from this sampling.

The observation error variance $\phi$ is estimated during the calibration. $\phi$ has an Inverse Gamma distribution ($IG)$ prior \mbox{~\cite{mebane2013PCCP}}. In principle, the $IG$ should be based on an \textit{a priori} assessment of error in the experiment. We have generated these $IG$ by considering the mode and the mean to be twice and four times as large as the largest standard deviation of the target parameters, respectively. The standard deviations used for bandgap, magnetic moment, and volume were 0.01 eV, 0.01 $\AA^3$, and 0.1 $\mu_B$ respectively. This results in $\phi_{E_g}\sim  IG(3, 0.08)$, $\phi_V \sim IG(3, 0.08)$, and $\phi_\mu \sim IG(3, 0.8)$.

\begin{table}
\centering
\caption{Convergence parameters. \textit{k}-grid shows the dimension of the selected grid for the Brillouin zone. E$_{cut}$ shows the cutoff of plane-wave expansion cutoff and it is presented in $eV$.}
\label{SI:tab:convergenc}

\begin{tabular}{l@{\hskip 1in}ll@{\hskip 1in}ll@{\hskip 1in}ll}
\hline \hline
Material  & LDA                 &                         & PBE                 &                           & PBEsol              &                         \\ \hline 
          & \textit{k}-grid              & E$_{cut}$ & \textit{k}-grid              & E$_{cut}$      & \textit{k}-grid              & E$_{cut}$      \\ \hline \hline
Fe        & 9$\times$9$\times$9 & 450 & 9$\times$9$\times$9 & 500   & 8$\times$8$\times$8 & 550 \\
Fe$_2$P   & 3$\times$3$\times$4 & 400 & 3$\times$3$\times$4 & 500   & 3$\times$3$\times$4 & 500 \\
Fe$_3$Ge  & 3$\times$3$\times$4 & 450 & 3$\times$3$\times$4 & 500   & 3$\times$3$\times$4 & 500 \\
SrFeO$_3$ & 8$\times$8$\times$8 & 900 & 8$\times$8$\times$8 & 900   & 8$\times$8$\times$8 & 900 \\
BaFeO$_3$ & 8$\times$8$\times$8 & 900 & 8$\times$8$\times$8 &  950 & 8$\times$8$\times$8 & 900 \\ \hline
\end{tabular}
\end{table}

\begin{figure}
\centering
  \includegraphics[width=\textwidth]{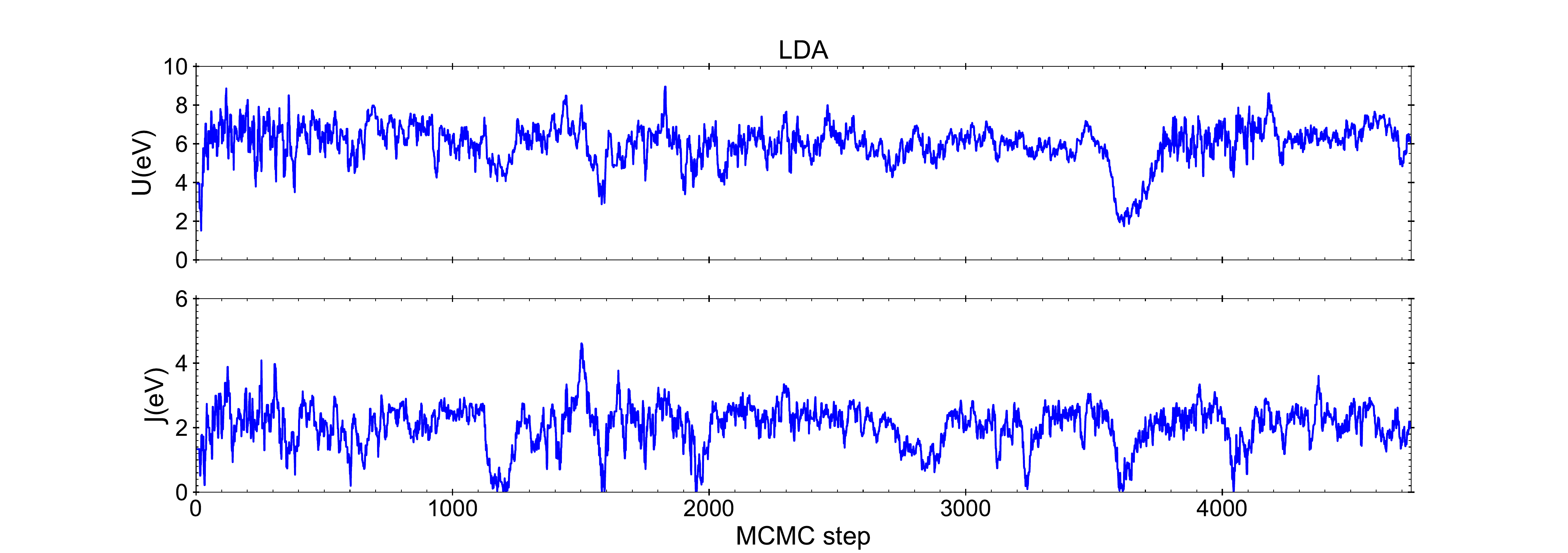}
  
  \includegraphics[width=\textwidth]{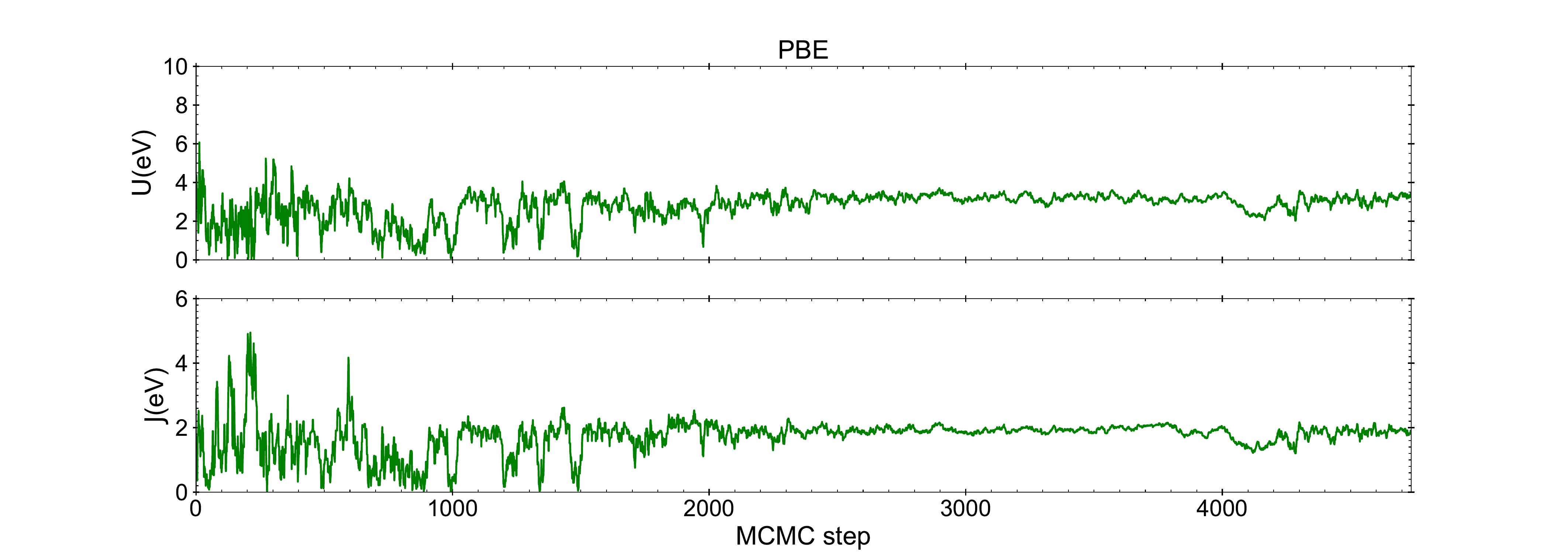}
  
  \includegraphics[width=\textwidth]{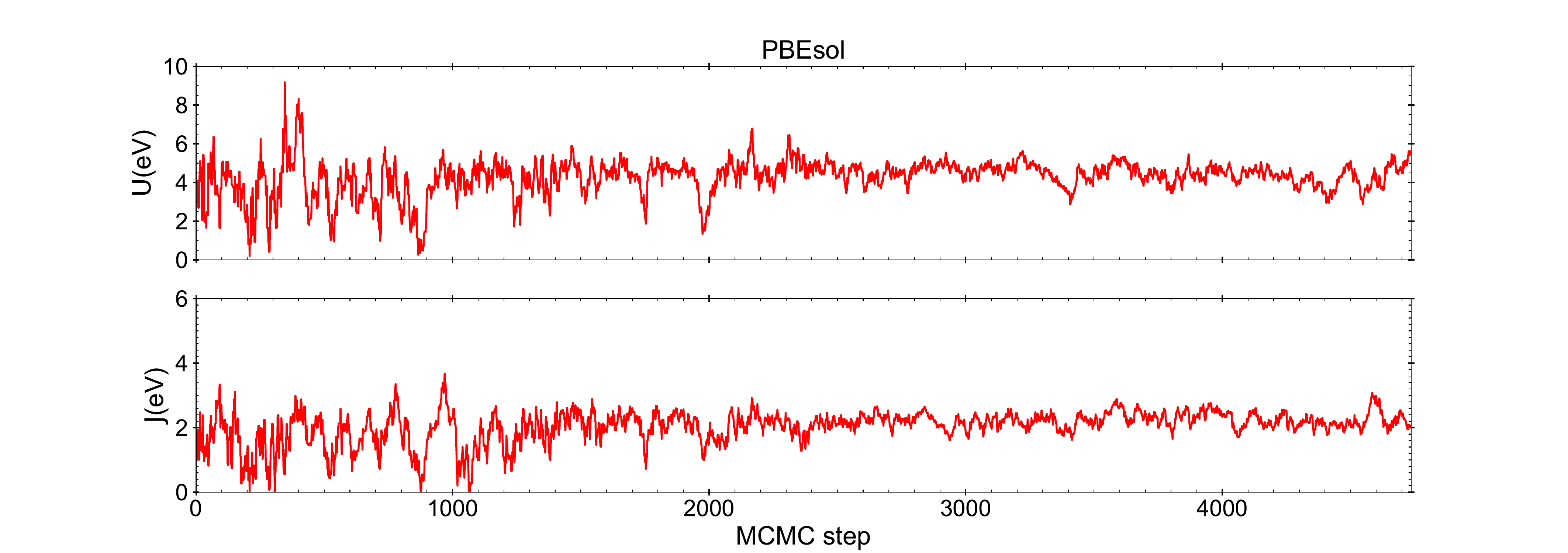}
  \caption{Trace of MCMC algorithm for LDA, PBE, and PBEsol XC functionals.}
  \label{SI:fig:trace}
\end{figure}

\begin{table}
\caption{Lattice parameter prediction evaluation for each XC functional and their corresponding Hubbard correction. Red, green, and white backgrounds depict underestimation, overestimation, and accurate, respectively. The accuracy criterion is 0.09 \r{A}. $\Delta$ represents the prediction error.}
\label{SI:tab:lattice_eval}
\resizebox{\textwidth}{!}{%
\begin{tabular}{lcccccccccccccc}
\hline \hline

\rotatebox[origin=c]{90}{Material} &
  \rotatebox[origin=c]{90}{Parameter} &
  \rotatebox[origin=c]{90}{Exp.} &
  \rotatebox[origin=c]{90}{LDA} &
  \rotatebox[origin=c]{90}{LDA+$U$} &
  \rotatebox[origin=c]{90}{PBE} &
  \rotatebox[origin=c]{90}{PBE+$U$} &
  \rotatebox[origin=c]{90}{PBEsol} &
  \rotatebox[origin=c]{90}{PBEsol+$U$} &
  \rotatebox[origin=c]{90}{$\Delta_{LDA}$} &
  \rotatebox[origin=c]{90}{$\Delta_{LDA+U}$} &
  \rotatebox[origin=c]{90}{$\Delta_{PBE}$} &
  \rotatebox[origin=c]{90}{$\Delta_{PBE+U}$} &
  \rotatebox[origin=c]{90}{$\Delta_{PBEsol}$} &
  \rotatebox[origin=c]{90}{$\Delta_{PBEsol+U}$} \\ \hline \hline
\ce{Fe} &
  a &
  2.87 &
  2.75 &
  2.83 &
  2.83 &
  2.84 &
  2.78 &
  2.85 &
  \cellcolor[HTML]{FFC7CE}{\color[HTML]{9C0006} -0.12} &
  -0.04 &
  -0.04 &
  -0.03 &
  \cellcolor[HTML]{FFC7CE}{\color[HTML]{9C0006} -0.09} &
  -0.02 \\
\ce{Fe2P} &
  a &
  5.87 &
  5.56 &
  5.88 &
  5.81 &
  5.91 &
  5.70 &
  5.90 &
  \cellcolor[HTML]{FFC7CE}{\color[HTML]{9C0006} -0.31} &
  0.01 &
  -0.06 &
  0.04 &
  \cellcolor[HTML]{FFC7CE}{\color[HTML]{9C0006} -0.17} &
  0.03 \\
\ce{Fe3Ge} &
  a &
  5.17 &
  4.95 &
  5.18 &
  5.14 &
  5.17 &
  5.15 &
  5.17 &
  \cellcolor[HTML]{FFC7CE}{\color[HTML]{9C0006} -0.22} &
  0.01 &
  -0.03 &
  0.00 &
  -0.02 &
  0.00 \\
\ce{BaFeO3} &
  a &
  3.97 &
  3.86 &
  3.90 &
  3.97 &
  3.98 &
  3.90 &
  3.91 &
  \cellcolor[HTML]{FFC7CE}{\color[HTML]{9C0006} -0.11} &
  -0.07 &
  0.00 &
  0.01 &
  -0.07 &
  -0.06 \\
\ce{SrFeO3} &
  a &
  3.85 &
  3.74 &
  3.78 &
  3.84 &
  3.85 &
  3.77 &
  3.79 &
  \cellcolor[HTML]{FFC7CE}{\color[HTML]{9C0006} -0.11} &
  -0.07 &
  -0.01 &
  0.00 &
  -0.08 &
  -0.06 \\
\ce{FeO} &
  a &
  4.31 &
  4.15 &
  4.20 &
  4.24 &
  4.27 &
  4.15 &
  4.22 &
  \cellcolor[HTML]{FFC7CE}{\color[HTML]{9C0006} -0.16} &
  \cellcolor[HTML]{FFC7CE}{\color[HTML]{9C0006} -0.11} &
  -0.07 &
  -0.04 &
  \cellcolor[HTML]{FFC7CE}{\color[HTML]{9C0006} -0.16} &
  \cellcolor[HTML]{FFC7CE}{\color[HTML]{9C0006} -0.09} \\
\ce{$\alpha$-Fe2O3} &
  a &
  5.03 &
  4.62 &
  4.95 &
  5.00 &
  5.05 &
  4.91 &
  5.00 &
  \cellcolor[HTML]{FFC7CE}{\color[HTML]{9C0006} -0.41} &
  -0.08 &
  -0.03 &
  0.02 &
  \cellcolor[HTML]{FFC7CE}{\color[HTML]{9C0006} -0.12} &
  -0.03 \\
\ce{AlFeB2} &
  a &
  2.92 &
  2.90 &
  2.87 &
  2.92 &
  2.92 &
  2.92 &
  2.92 &
  -0.02 &
  -0.05 &
  0.00 &
  0.00 &
  0.00 &
  0.00 \\
\ce{Fe5PB2} &
  a &
  5.49 &
  5.45 &
  5.45 &
  5.44 &
  5.51 &
  5.35 &
  5.48 &
  -0.04 &
  -0.04 &
  -0.05 &
  0.02 &
  \cellcolor[HTML]{FFC7CE}{\color[HTML]{9C0006} -0.14} &
  -0.01 \\
\ce{Fe5SiB2} &
  a &
  5.55 &
  5.45 &
  5.45 &
  5.50 &
  5.54 &
  5.43 &
  5.51 &
  \cellcolor[HTML]{FFC7CE}{\color[HTML]{9C0006} -0.10} &
  \cellcolor[HTML]{FFC7CE}{\color[HTML]{9C0006} -0.10} &
  -0.05 &
  -0.01 &
  \cellcolor[HTML]{FFC7CE}{\color[HTML]{9C0006} -0.12} &
  -0.04 \\
\ce{AlFeB2} &
  b &
  11.03 &
  11.13 &
  10.84 &
  11.01 &
  11.01 &
  11.01 &
  11.01 &
  \cellcolor[HTML]{C6EFCE}{\color[HTML]{006100} 0.10} &
  \cellcolor[HTML]{FFC7CE}{\color[HTML]{9C0006} -0.19} &
  -0.02 &
  -0.02 &
  -0.02 &
  -0.02 \\
\ce{Fe2P} &
  c &
  3.46 &
  3.42 &
  3.32 &
  3.41 &
  3.38 &
  3.40 &
  3.36 &
  -0.04 &
  \cellcolor[HTML]{FFC7CE}{\color[HTML]{9C0006} -0.14} &
  -0.05 &
  -0.08 &
  -0.06 &
  \cellcolor[HTML]{FFC7CE}{\color[HTML]{9C0006} -0.10} \\
\ce{Fe3Ge} &
  c &
  4.22 &
  4.03 &
  4.17 &
  4.20 &
  4.21 &
  4.22 &
  4.28 &
  \cellcolor[HTML]{FFC7CE}{\color[HTML]{9C0006} -0.19} &
  -0.05 &
  -0.02 &
  -0.01 &
  0.00 &
  0.06 \\
\ce{$\alpha$-Fe2O3} &
  c &
  13.75 &
  13.31 &
  13.60 &
  13.86 &
  13.91 &
  13.66 &
  13.73 &
  \cellcolor[HTML]{FFC7CE}{\color[HTML]{9C0006} -0.44} &
  \cellcolor[HTML]{FFC7CE}{\color[HTML]{9C0006} -0.15} &
  \cellcolor[HTML]{C6EFCE}{\color[HTML]{006100} 0.11} &
  \cellcolor[HTML]{C6EFCE}{\color[HTML]{006100} 0.16} &
  \cellcolor[HTML]{FFC7CE}{\color[HTML]{9C0006} -0.09} &
  -0.02 \\
\ce{AlFeB2} &
  c &
  2.87 &
  2.64 &
  2.85 &
  2.86 &
  2.86 &
  2.86 &
  2.86 &
  \cellcolor[HTML]{FFC7CE}{\color[HTML]{9C0006} -0.23} &
  -0.02 &
  -0.01 &
  -0.01 &
  -0.01 &
  -0.01 \\
\ce{Fe5PB2} &
  c &
  10.35 &
  10.31 &
  10.31 &
  10.34 &
  10.39 &
  10.18 &
  10.26 &
  -0.04 &
  -0.04 &
  -0.01 &
  0.04 &
  \cellcolor[HTML]{FFC7CE}{\color[HTML]{9C0006} -0.17} &
  \cellcolor[HTML]{FFC7CE}{\color[HTML]{9C0006} -0.09} \\
\ce{Fe5SiB2} &
  c &
  10.34 &
  10.31 &
  10.31 &
  10.33 &
  10.42 &
  10.12 &
  10.27 &
  -0.03 &
  -0.03 &
  -0.01 &
  0.08 &
  \cellcolor[HTML]{FFC7CE}{\color[HTML]{9C0006} -0.22} &
  -0.07 \\ \hline
\end{tabular}%
}
\end{table}

\begin{table}
\centering
\caption{Magnetic moment prediction evaluation for each XC functional and their corresponding Hubbard correction. Red, green, and white backgrounds depict underestimation, overestimation, and accurate, respectively. The accuracy criterion is 0.2 $\mu_B$. $\Delta$ represents the prediction error.}
\label{SI:tab:magmom_eval}
\begin{tabular}{l@{\hskip 0.25in}c@{\hskip 0.2in}cccccc@{\hskip 0.2in}cccccc}
 \hline \hline
\rowcolor[HTML]{FFFFFF}
\rotatebox[origin=c]{90}{Material} &
  \rotatebox[origin=c]{90}{Exp.} &
  \rotatebox[origin=c]{90}{LDA} &
  \rotatebox[origin=c]{90}{LDA+$U$} &
  \rotatebox[origin=c]{90}{PBE} &
  \rotatebox[origin=c]{90}{PBE+$U$}&
  \rotatebox[origin=c]{90}{PBEsol} &
  \rotatebox[origin=c]{90}{PBEsol+$U$} &
  \rotatebox[origin=c]{90}{$\Delta_{LDA}$} &
  \rotatebox[origin=c]{90}{$\Delta_{LDA+U}$} &
  \rotatebox[origin=c]{90}{$\Delta_{PBE}$} &
  \rotatebox[origin=c]{90}{$\Delta_{PBE+U}$} &
  \rotatebox[origin=c]{90}{$\Delta_{PBEsol}$} &
  \rotatebox[origin=c]{90}{$\Delta_{PBEsol+U}$} \\ \hline \hline
\ce{Fe} &
  2.22 &
  1.95 &
  2.73 &
  2.19 &
  2.09 &
  2.12 &
  2.71 &
  \cellcolor[HTML]{FFC7CE}{\color[HTML]{9C0006} -0.27} &
  \cellcolor[HTML]{C6EFCE}{\color[HTML]{006100} 0.51} &
  -0.03 &
  -0.13 &
  \cellcolor[HTML]{FFC7CE}{\color[HTML]{9C0006} -0.10} &
  \cellcolor[HTML]{C6EFCE}{\color[HTML]{006100} 0.49} \\
\ce{Fe2P} &
  1.91 &
  1.11 &
  2.26 &
  2.25 &
  2.09 &
  2.03 &
  2.23 &
  \cellcolor[HTML]{FFC7CE}{\color[HTML]{9C0006} -0.80} &
  \cellcolor[HTML]{C6EFCE}{\color[HTML]{006100} 0.35} &
  \cellcolor[HTML]{C6EFCE}{\color[HTML]{006100} 0.34} &
  0.18 &
  \cellcolor[HTML]{FFC7CE}{\color[HTML]{9C0006} 0.12} &
  \cellcolor[HTML]{C6EFCE}{\color[HTML]{006100} 0.32} \\
\ce{Fe3Ge} &
  2.00 &
  1.25 &
  2.75 &
  2.18 &
  2.37 &
  2.17 &
  2.66 &
  \cellcolor[HTML]{FFC7CE}{\color[HTML]{9C0006} -0.75} &
  \cellcolor[HTML]{C6EFCE}{\color[HTML]{006100} 0.75} &
  0.18 &
  \cellcolor[HTML]{C6EFCE}{\color[HTML]{006100} 0.37} &
  0.17 &
  \cellcolor[HTML]{C6EFCE}{\color[HTML]{006100} 0.66} \\
\ce{BaFeO3} &
  3.50 &
  2.64 &
  3.56 &
  3.02 &
  3.37 &
  2.88 &
  3.45 &
  \cellcolor[HTML]{FFC7CE}{\color[HTML]{9C0006} -0.86} &
  0.06 &
  \cellcolor[HTML]{FFC7CE}{\color[HTML]{9C0006} -0.48} &
  -0.13 &
  \cellcolor[HTML]{FFC7CE}{\color[HTML]{9C0006} -0.62} &
  -0.05 \\
\ce{SrFeO3} &
  3.10 &
  2.51 &
  3.49 &
  2.87 &
  3.15 &
  2.71 &
  3.36 &
  \cellcolor[HTML]{FFC7CE}{\color[HTML]{9C0006} -0.59} &
  \cellcolor[HTML]{C6EFCE}{\color[HTML]{006100} 0.39} &
  \cellcolor[HTML]{FFC7CE}{\color[HTML]{9C0006} -0.23} &
  0.05 &
  \cellcolor[HTML]{FFC7CE}{\color[HTML]{9C0006} -0.39} &
  \cellcolor[HTML]{C6EFCE}{\color[HTML]{006100} 0.26} \\
\ce{FeO} &
  3.32 &
  3.30 &
  0.12 &
  3.40 &
  3.51 &
  3.29 &
  3.55 &
  \cellcolor[HTML]{FFC7CE}{\color[HTML]{9C0006} -0.02} &
  \cellcolor[HTML]{FFC7CE}{\color[HTML]{9C0006} -3.20} &
  0.08 &
  0.19 &
  \cellcolor[HTML]{FFC7CE}{\color[HTML]{9C0006} -0.03} &
  \cellcolor[HTML]{C6EFCE}{\color[HTML]{006100} 0.23} \\
\ce{$\alpha$-Fe2O3} &
  4.90 &
  1.11 &
  4.00 &
  3.55 &
  3.85 &
  3.36 &
  3.95 &
  \cellcolor[HTML]{FFC7CE}{\color[HTML]{9C0006} -3.79} &
  \cellcolor[HTML]{FFC7CE}{\color[HTML]{9C0006} -0.90} &
  \cellcolor[HTML]{FFC7CE}{\color[HTML]{9C0006} -1.35} &
  \cellcolor[HTML]{FFC7CE}{\color[HTML]{9C0006} -1.05} &
  \cellcolor[HTML]{FFC7CE}{\color[HTML]{9C0006} -1.54} &
  \cellcolor[HTML]{FFC7CE}{\color[HTML]{9C0006} -0.95} \\
\ce{AlFeB2} &
  1.21 &
  0.00 &
  1.64 &
  1.40 &
  1.52 &
  1.37 &
  1.57 &
  \cellcolor[HTML]{FFC7CE}{\color[HTML]{9C0006} -1.21} &
  \cellcolor[HTML]{C6EFCE}{\color[HTML]{006100} 0.43} &
  0.19 &
  \cellcolor[HTML]{C6EFCE}{\color[HTML]{006100} 0.31} &
  0.16 &
  \cellcolor[HTML]{C6EFCE}{\color[HTML]{006100} 0.36} \\
\ce{Fe5PB2} &
  1.73 &
  1.43 &
  2.21 &
  1.79 &
  1.99 &
  1.55 &
  2.11 &
  \cellcolor[HTML]{FFC7CE}{\color[HTML]{9C0006} -0.30} &
  \cellcolor[HTML]{C6EFCE}{\color[HTML]{006100} 0.48} &
  0.06 &
  \cellcolor[HTML]{C6EFCE}{\color[HTML]{006100} 0.26} &
  \cellcolor[HTML]{FFC7CE}{\color[HTML]{9C0006} -0.18} &
  \cellcolor[HTML]{C6EFCE}{\color[HTML]{006100} 0.38} \\
\ce{Fe5SiB2} &
  1.83 &
  1.48 &
  2.11 &
  1.84 &
  1.98 &
  1.61 &
  2.04 &
  \cellcolor[HTML]{FFC7CE}{\color[HTML]{9C0006} -0.35} &
  \cellcolor[HTML]{C6EFCE}{\color[HTML]{006100} 0.28} &
  0.01 &
  0.15 &
  \cellcolor[HTML]{FFC7CE}{\color[HTML]{9C0006} -0.22} &
  \cellcolor[HTML]{C6EFCE}{\color[HTML]{006100} 0.21} \\\hline
\end{tabular}
\end{table}

\begin{table}
\centering
\caption{Root mean square error (RMSE) and mean absolute error (MAE) for prediction using DFT and DFT+$U$. Note, after including the Hubbard correction, predictions for volume and magnetic moment ($V$ and $\mu$) improve. Green represents improvement in predictions.}
\label{SI:tab:rmse_mae}
\begin{tabular}{l@{\hskip 0.6in}l@{\hskip 0.4in}
cc@{\hskip 0.3in}
cc@{\hskip 0.3in}
cc@{\hskip 0.4in}
cc}
\hline \hline
 &
   Target Property &
  \rotatebox[origin=c]{90}{LDA} &
  \rotatebox[origin=c]{90}{LDA+$U$} &
  \rotatebox[origin=c]{90}{PBE} &
  \rotatebox[origin=c]{90}{PBE+$U$} &
  \rotatebox[origin=c]{90}{PBEsol} &
  \rotatebox[origin=c]{90}{PBEsol+$U$} &
  \rotatebox[origin=c]{90}{DFT} &
  \rotatebox[origin=c]{90}{DFT+$U$} \\ \hline \hline
RSME & Volume $(\AA^3)$     &
22.34 & 
 \cellcolor[HTML]{C6EFCE}{\color[HTML]{006100} 10.50} &
 8.77 & 
 \cellcolor[HTML]{C6EFCE}{\color[HTML]{006100} 7.54} & 
 14.11 &
 \cellcolor[HTML]{C6EFCE}{\color[HTML]{006100}7.91 }&
 16.07 & 
 \cellcolor[HTML]{C6EFCE}{\color[HTML]{006100} 8.75} \\
     & Mag. Mom.  $(\mu_B)$ &
     1.36  &
     \cellcolor[HTML]{C6EFCE}{\color[HTML]{006100} 1.12}  &
     0.48 & 
     \cellcolor[HTML]{C6EFCE}{\color[HTML]{006100} 0.39} &
     0.55  &
     \cellcolor[HTML]{C6EFCE}{\color[HTML]{006100} 0.46} &
     0.89  &
     \cellcolor[HTML]{C6EFCE}{\color[HTML]{006100} 0.74} \\ \hline
MAE  & Volume $(\AA^3)$     &
15.70 &
\cellcolor[HTML]{C6EFCE}{\color[HTML]{006100} 8.61}  &
5.55 &
\cellcolor[HTML]{C6EFCE}{\color[HTML]{006100} 4.69} &
11.47 &
\cellcolor[HTML]{C6EFCE}{\color[HTML]{006100} 5.69} &
10.91 &
\cellcolor[HTML]{C6EFCE}{\color[HTML]{006100} 6.33} \\
     & Mag. Mom.  $(\mu_B)$ &
     0.89  &
     \cellcolor[HTML]{C6EFCE}{\color[HTML]{006100} 0.74}  &
     0.30 & 
     \cellcolor[HTML]{C6EFCE}{\color[HTML]{006100} 0.28} &
     0.35  &
     \cellcolor[HTML]{C6EFCE}{\color[HTML]{006100} 0.39} &
     0.51  & 
     \cellcolor[HTML]{C6EFCE}{\color[HTML]{006100} 0.47} \\ \hline \hline
\end{tabular}

\end{table}

\begin{table}
\caption{Structural, electronic, and magnetic properties of \ce{BaFe2As2}. Values outside \textcolor{blue}{(inside)} parenthesis are from simulations using DFT \textcolor{blue}{(DFT+$U$)}. The DFT+$U$ calculations were performed using the mean values of $U$ and $J$ from the distributions. Letters a, b, and c represent the lattice parameters. MP represents the final magnetic phase. 
Volume and magnetic moment are expressed in units of \r{A}$^3$ and Bohr magneton ($\mu_B$), respectively.}
\label{SI:tab:BaFe2As2}
\resizebox{\textwidth}{!}{
\begin{tabular}{llllllll}
\hline \hline
Material &
  XC &
  a &
  b &
  c &
  Volume &
  Mag. Mom. &
  MP \\ \hline \hline
\ce{BaFe2As2} &
  Experiment &
  5.61 \textsuperscript{\emph{a}} &
  5.57 \textsuperscript{\emph{a}} &
  12.95 \textsuperscript{\emph{a}} &
  405.14 \textsuperscript{\emph{a}} &
  0.40, 0.5, 0.87 \textsuperscript{\emph{a},\emph{b},\emph{c}} &
  AFM \textsuperscript{\emph{a}} \\ \hline
Ferromagnet &
  LDA \textcolor{blue}{(+$U$)} &
  5.48 \textcolor{blue}{(5.54)} &
  5.48 \textcolor{blue}{(5.54)} &
  12.30 \textcolor{blue}{(13.32)} &
  369.52 \textcolor{blue}{(408.79)} &
  0.00 \textcolor{blue}{(1.74)} &
  NM \textcolor{blue}{(FM)} \\
 &
  PBE \textcolor{blue}{(+$U$)} &
  5.60 \textcolor{blue}{(5.64)} &
  5.60 \textcolor{blue}{(5.63)} &  
   12.60 \textcolor{blue}{(13.49)} &
   395.86 \textcolor{blue}{(427.83)} &
   0.00 \textcolor{blue}{(1.43)} &
  NM \textcolor{blue}{(FM)} \\
 &
  PBEsol \textcolor{blue}{(+$U$)} &
   5.53 \textcolor{blue}{(5.56)} &
   5.53 \textcolor{blue}{(5.55)} &
   12.34 \textcolor{blue}{(13.37)} &
   377.35 \textcolor{blue}{(411.56)} &
   0.00 \textcolor{blue}{(1.56)} &
  NM \textcolor{blue}{(FM)} \\ \hline

Antiferromagnet &
  LDA \textcolor{blue}{(+$U$)} &
  5.48 \textcolor{blue}{(5.82)} &
  5.48 \textcolor{blue}{(5.40)} &
  12.30 \textcolor{blue}{(13.19)} &
  369.50 \textcolor{blue}{(415.03)} &
  0.00 \textcolor{blue}{(2.89)} &
  NM \textcolor{blue}{(AFM)} \\
 stripe &
  PBE \textcolor{blue}{(+$U$)} &
  5.68 \textcolor{blue}{(5.93)} &
  5.58 \textcolor{blue}{(5.58)} &
  12.88 \textcolor{blue}{(13.16)} &
  408.39 \textcolor{blue}{(435.75)} &
  1.90 \textcolor{blue}{(2.73)} &
  AFM \textcolor{blue}{(AFM)} \\
 &
  PBEsol \textcolor{blue}{(+$U$)} &
  5.54 \textcolor{blue}{(5.85)} &
  5.54 \textcolor{blue}{(5.85)} &
  12.65 \textcolor{blue}{(13.09)} &
  380.99 \textcolor{blue}{(420.13)} &
  1.11 \textcolor{blue}{(2.86)} &
  AFM \textcolor{blue}{(AFM)} \\ \hline 
Antiferromagnet &
  LDA \textcolor{blue}{(+$U$)} &
   5.48 \textcolor{blue}{(5.88)} &
   5.48 \textcolor{blue}{(5.29)} &
   12.30 \textcolor{blue}{(13.42)} &
   369.53 \textcolor{blue}{(417.64)} &
   0.00 \textcolor{blue}{(2.50)} &
  NM \textcolor{blue}{(AFM)} \\
checkerboard  &
  PBE \textcolor{blue}{(+$U$)} &
  5.63 \textcolor{blue}{(6.01)} &
  5.63 \textcolor{blue}{(5.47)} &
  12.70 \textcolor{blue}{(13.36)} &
  403.05 \textcolor{blue}{(439.60)} &
  1.38 \textcolor{blue}{(2.17)} &
  AFM \textcolor{blue}{(AFM)} \\
 &
  PBEsol \textcolor{blue}{(+$U$)} &
  5.53 \textcolor{blue}{(5.91)} &
  5.53 \textcolor{blue}{(5.35)} &
  12.34 \textcolor{blue}{(13.36)} &
  377.36 \textcolor{blue}{(422.91)} &
  0.00 \textcolor{blue}{(2.39)} &
  NM \textcolor{blue}{(AFM)} \\ \hline \hline
\end{tabular}}
\textsuperscript{\emph{a}}
Ref. ~\citenum{BaFeAs-crystal};
\textsuperscript{\emph{b}}
Ref. ~\citenum{BaFeAs-magmom1}; 
\textsuperscript{\emph{c}}
Ref. ~\citenum{BaFeAs-magmom2}; 
\end{table}   

\begin{figure}
\centering
  \includegraphics[width=\textwidth]{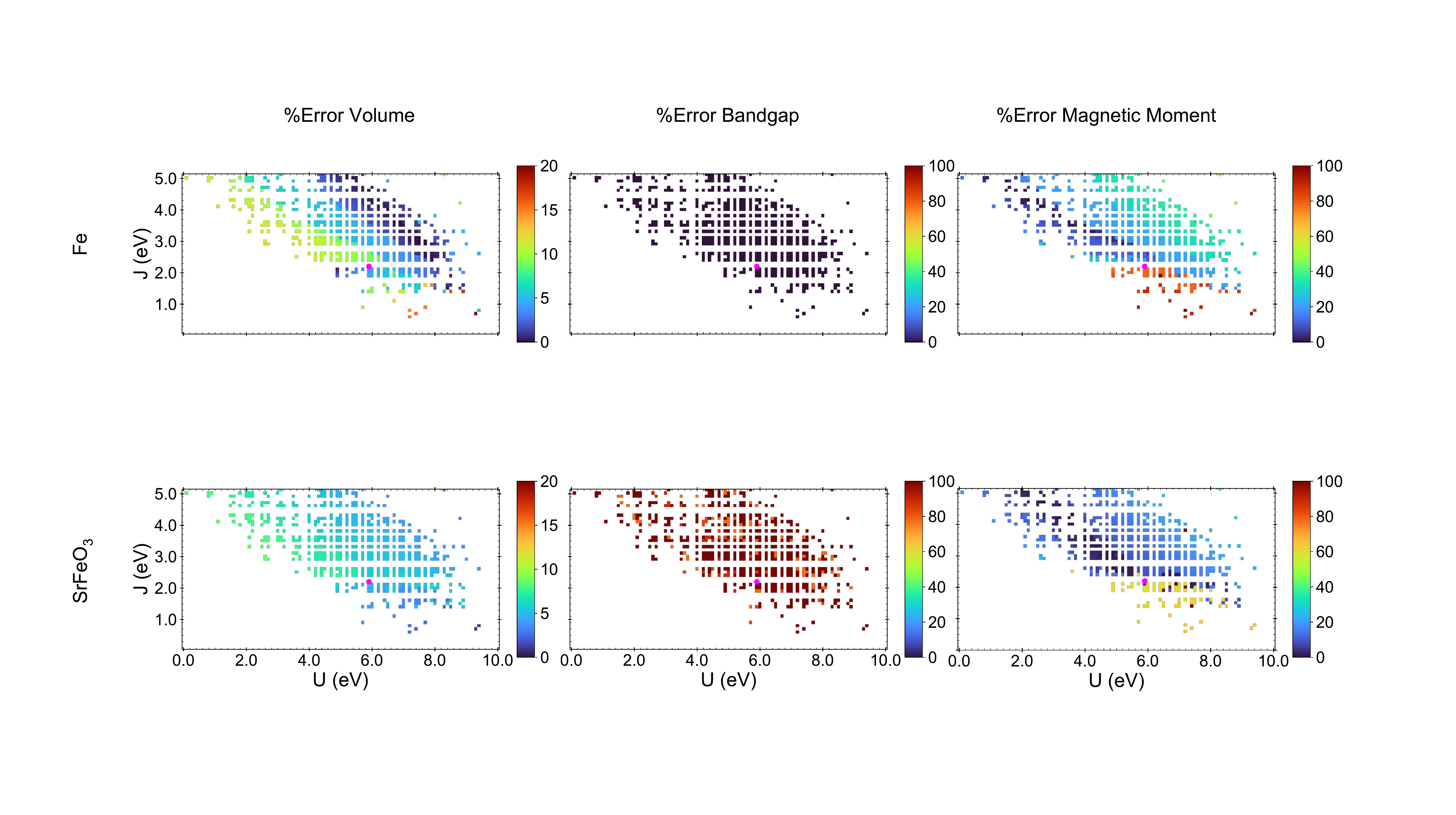}

  \caption{LDA+$U$ prediction percentage error heat maps of $U$ and $J$ for \ce{Fe} and \ce{SrFeO3} for volume, bandgap, and magnetic moment.}
  \label{SI:fig:lda_heat_map}
\end{figure}

\begin{table}
\centering
\caption{$U$ and $J$ values extracted from MCMC sampling with and without bandgap as a target parameter. $\mu$, V, and $E_g$ represent magnetic moment, volume, and bandgap, respectively.}
\label{SI:tab:noBG}
\begin{tabular}{l@{\hskip 1.1in}c@{\hskip 0.6in}c@{\hskip 1.1in}c@{\hskip 0.6in}c}
\hline \hline 
XC                & LDA+$U$            & LDA+$U$     & PBE+$U$            & PBE+$U$     \\
Target properties & $\mu$,  $V$, $E_g$ & $\mu$,  $V$ & $\mu$,  $V$, $E_g$ & $\mu$,  $V$ \\ \hline \hline
$U$ (eV)            & 5.9                & 6.4         & 3.1                & 3.1         \\
$J$ (eV)            & 2.1                & 1.9         & 1.9                & 1.8        \\
\hline \hline
\end{tabular}
\end{table}

\begin{figure}
\centering
  \vspace*{-6cm}\includegraphics[width=16cm]{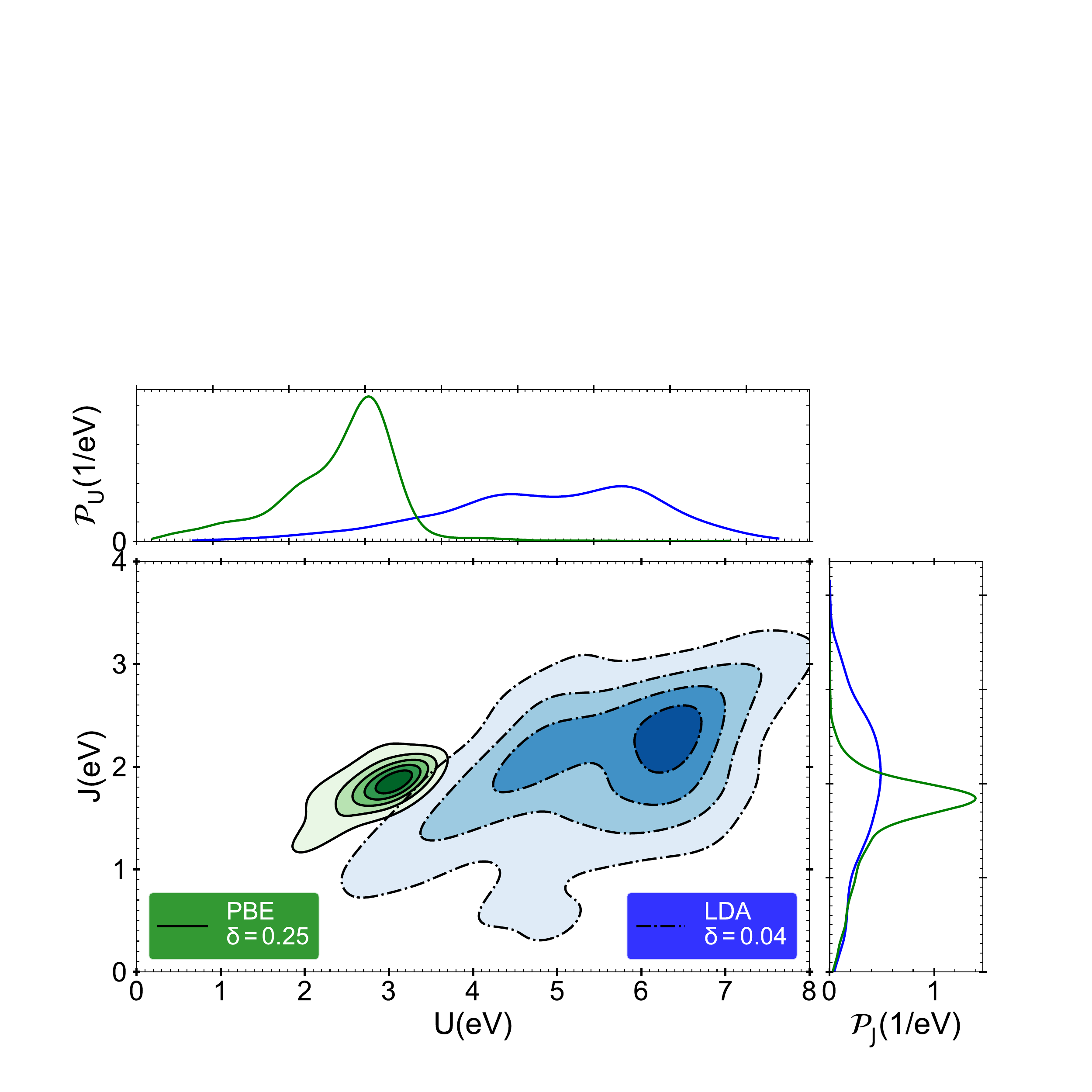}

  \caption{$U$ and $J$ values extracted from MCMC sampling without bandgap in the target properties. Target properties in this sampling are volume and magnetic moment. The density was estimated using a gaussian kernel density estimation (KDE). The bandwidth was selected using the Scott approach. Each KDE is separately normalized to one. $\delta$ is the step between contour lines. Top left shows the probability density function of accepted $U$ parameters. Bottom left shows the joint probability density function of accepted $U$ and $J$. Bottom right shows the probability density function of accepted $J$ parameters.}
  \label{SI:fig:distribution_noBG}
\end{figure}

\begin{table}
\centering
\caption{RMSE and MAE of DFT+$U$ with $U$ and $J$ values extracted from MCMC samplings with and without bandgap included in the target parameters.}
\label{SI:tab:noBG_rmse}
\begin{tabular}{l@{\hskip 1.1in}c@{\hskip 0.6in}c@{\hskip 1.1in}c@{\hskip 0.6in}c}
\hline \hline 
  & RMSE              & RMSE       & MAE               & MAE        \\
 Target properties  & $\mu$, $V$, $E_g$ & $\mu$, $V$ & $\mu$, $V$, $E_g$ & $\mu$, $V$ \\
\hline \hline 
Volume   ($\AA^3$)  & 8.75              & \cellcolor[HTML]{C6EFCE}{\color[HTML]{006100}6.68}       & 6.33              & \cellcolor[HTML]{C6EFCE}{\color[HTML]{006100}3.98}       \\
Mag. Mom. ($\mu_B$) & 0.74              & \cellcolor[HTML]{C6EFCE}{\color[HTML]{006100}0.39}       & 0.47              & \cellcolor[HTML]{C6EFCE}{\color[HTML]{006100}0.26}     \\
\hline \hline
\end{tabular}
\end{table}


\begin{thebibliography}{100}
\expandafter\ifx\csname url\endcsname\relax
  \def\url#1{\texttt{#1}}\fi
\expandafter\ifx\csname urlprefix\endcsname\relax\def\urlprefix{URL }\fi
\providecommand{\bibinfo}[2]{#2}
\providecommand{\eprint}[2][]{\url{#2}}

\bibitem{hehenberg-kohn}
\bibinfo{author}{Hohenberg, P.} \& \bibinfo{author}{Kohn, W.}
\newblock \bibinfo{title}{Inhomogeneous Electron Gas}.
\newblock \emph{\bibinfo{journal}{Phys. Rev.}} \textbf{\bibinfo{volume}{136}},
  \bibinfo{pages}{B864--B871} (\bibinfo{year}{1964}).
\newblock \urlprefix\url{https://link.aps.org/doi/10.1103/PhysRev.136.B864}.

\bibitem{kohn1965self}
\bibinfo{author}{Kohn, W.} \& \bibinfo{author}{Sham, L.~J.}
\newblock \bibinfo{title}{Self-Consistent Equations Including Exchange and
  Correlation Effects}.
\newblock \emph{\bibinfo{journal}{Phys. Rev.}} \textbf{\bibinfo{volume}{140}},
  \bibinfo{pages}{A1133--A1138} (\bibinfo{year}{1965}).
\newblock \urlprefix\url{https://link.aps.org/doi/10.1103/PhysRev.140.A1133}.

\bibitem{kohn1996density}
\bibinfo{author}{Kohn, W.}, \bibinfo{author}{Becke, A.~D.} \&
  \bibinfo{author}{Parr, R.~G.}
\newblock \bibinfo{title}{Density Functional Theory of Electronic Structure}.
\newblock \emph{\bibinfo{journal}{J. Phys. Chem.}}
  \textbf{\bibinfo{volume}{100}}, \bibinfo{pages}{12974--12980}
  (\bibinfo{year}{1996}).
\newblock \urlprefix\url{https://doi.org/10.1021/jp960669l}.

\bibitem{sholl2011density-DFT}
\bibinfo{author}{Steckel, J.~A.} \& \bibinfo{author}{Sholl, D.}
\newblock \emph{\bibinfo{title}{Density Functional Theory}}
  (\bibinfo{publisher}{John Wiley \& Sons, Ltd}, \bibinfo{address}{Hoboken},
  \bibinfo{year}{2009}).
\newblock
  \urlprefix\url{https://onlinelibrary.wiley.com/doi/abs/10.1002/9780470447710}.

\bibitem{fiolhais2003primer-DFT}
\bibinfo{author}{Fiolhais, C.}, \bibinfo{author}{Nogueira, F.} \&
  \bibinfo{author}{Marques, M.~A.}
\newblock \emph{\bibinfo{title}{A primer in density functional theory}}, vol.
  \bibinfo{volume}{620} (\bibinfo{publisher}{Springer Berlin Heidelberg},
  \bibinfo{address}{Berlin Heidelberg}, \bibinfo{year}{2003}).
\newblock \urlprefix\url{https://doi.org/10.1007/3-540-37072-2}.

\bibitem{parr1980density-DFT}
\bibinfo{author}{Parr, R.~G.}
\newblock \bibinfo{title}{Density Functional Theory of Atoms and Molecules}.
\newblock In \bibinfo{editor}{Fukui, K.} \& \bibinfo{editor}{Pullman, B.}
  (eds.) \emph{\bibinfo{booktitle}{Horizons of Quantum Chemistry}},
  \bibinfo{pages}{5--15} (\bibinfo{publisher}{Springer},
  \bibinfo{address}{Dordrecht}, \bibinfo{year}{1980}).

\bibitem{b88}
\bibinfo{author}{Becke, A.~D.}
\newblock \bibinfo{title}{Density-functional exchange-energy approximation with
  correct asymptotic behavior}.
\newblock \emph{\bibinfo{journal}{Phys. Rev. A}} \textbf{\bibinfo{volume}{38}},
  \bibinfo{pages}{3098--3100} (\bibinfo{year}{1988}).
\newblock \urlprefix\url{https://link.aps.org/doi/10.1103/PhysRevA.38.3098}.

\bibitem{pw91}
\bibinfo{author}{Perdew, J.~P.} \emph{et~al.}
\newblock \bibinfo{title}{Atoms, molecules, solids, and surfaces: Applications
  of the generalized gradient approximation for exchange and correlation}.
\newblock \emph{\bibinfo{journal}{Phys. Rev. B}} \textbf{\bibinfo{volume}{46}},
  \bibinfo{pages}{6671--6687} (\bibinfo{year}{1992}).
\newblock \urlprefix\url{https://link.aps.org/doi/10.1103/PhysRevB.46.6671}.

\bibitem{lyp}
\bibinfo{author}{Lee, C.}, \bibinfo{author}{Yang, W.} \& \bibinfo{author}{Parr,
  R.~G.}
\newblock \bibinfo{title}{Development of the Colle-Salvetti correlation-energy
  formula into a functional of the electron density}.
\newblock \emph{\bibinfo{journal}{Phys. Rev. B}} \textbf{\bibinfo{volume}{37}},
  \bibinfo{pages}{785--789} (\bibinfo{year}{1988}).
\newblock \urlprefix\url{https://link.aps.org/doi/10.1103/PhysRevB.37.785}.

\bibitem{PBE-PRB}
\bibinfo{author}{Perdew, J.~P.}, \bibinfo{author}{Burke, K.} \&
  \bibinfo{author}{Wang, Y.}
\newblock \bibinfo{title}{Generalized gradient approximation for the
  exchange-correlation hole of a many-electron system}.
\newblock \emph{\bibinfo{journal}{Phys. Rev. B}} \textbf{\bibinfo{volume}{54}},
  \bibinfo{pages}{16533--16539} (\bibinfo{year}{1996}).
\newblock \urlprefix\url{https://link.aps.org/doi/10.1103/PhysRevB.54.16533}.

\bibitem{PBEsol}
\bibinfo{author}{Csonka, G.~I.} \emph{et~al.}
\newblock \bibinfo{title}{Assessing the performance of recent density
  functionals for bulk solids}.
\newblock \emph{\bibinfo{journal}{Phys. Rev. B}} \textbf{\bibinfo{volume}{79}},
  \bibinfo{pages}{155107} (\bibinfo{year}{2009}).
\newblock \urlprefix\url{https://link.aps.org/doi/10.1103/PhysRevB.79.155107}.

\bibitem{SCAN-sun2015strongly}
\bibinfo{author}{Sun, J.}, \bibinfo{author}{Ruzsinszky, A.} \&
  \bibinfo{author}{Perdew, J.~P.}
\newblock \bibinfo{title}{Strongly Constrained and Appropriately Normed
  Semilocal Density Functional}.
\newblock \emph{\bibinfo{journal}{Phys. Rev. Lett.}}
  \textbf{\bibinfo{volume}{115}}, \bibinfo{pages}{036402}
  (\bibinfo{year}{2015}).
\newblock
  \urlprefix\url{https://link.aps.org/doi/10.1103/PhysRevLett.115.036402}.

\bibitem{B3LYP1}
\bibinfo{author}{Kim, K.} \& \bibinfo{author}{Jordan, K.}
\newblock \bibinfo{title}{Comparison of density functional and MP2 calculations
  on the water monomer and dimer}.
\newblock \emph{\bibinfo{journal}{J. Phys. Chem.}}
  \textbf{\bibinfo{volume}{98}}, \bibinfo{pages}{10089--10094}
  (\bibinfo{year}{1994}).
\newblock \urlprefix\url{https://doi.org/10.1021/j100091a024}.

\bibitem{MN15-L}
\bibinfo{author}{Sun, J.}, \bibinfo{author}{Xiao, B.} \&
  \bibinfo{author}{Ruzsinszky, A.}
\newblock \bibinfo{title}{Communication: Effect of the orbital-overlap
  dependence in the meta generalized gradient approximation}.
\newblock \emph{\bibinfo{journal}{J. Chem. Phys.}}
  \textbf{\bibinfo{volume}{137}}, \bibinfo{pages}{051101}
  (\bibinfo{year}{2012}).
\newblock \urlprefix\url{https://doi.org/10.1063/1.4742312}.

\bibitem{XC-Progress}
\bibinfo{author}{Scuseria, G.~E.} \& \bibinfo{author}{Staroverov, V.~N.}
\newblock \bibinfo{title}{Chapter 24 - Progress in the development of
  exchange-correlation functionals}.
\newblock In \bibinfo{editor}{Dykstra, C.~E.}, \bibinfo{editor}{Frenking, G.},
  \bibinfo{editor}{Kim, K.~S.} \& \bibinfo{editor}{Scuseria, G.~E.} (eds.)
  \emph{\bibinfo{booktitle}{Theory and Applications of Computational
  Chemistry}}, \bibinfo{pages}{669--724} (\bibinfo{publisher}{Elsevier},
  \bibinfo{address}{Amsterdam}, \bibinfo{year}{2005}).
\newblock
  \urlprefix\url{https://www.sciencedirect.com/science/article/pii/B9780444517197500676}.

\bibitem{DFTthirtyyears}
\bibinfo{author}{Mardirossian, N.} \& \bibinfo{author}{Head-Gordon, M.}
\newblock \bibinfo{title}{Thirty years of density functional theory in
  computational chemistry: an overview and extensive assessment of 200 density
  functionals}.
\newblock \emph{\bibinfo{journal}{Mol. Phys.}} \textbf{\bibinfo{volume}{115}},
  \bibinfo{pages}{2315--2372} (\bibinfo{year}{2017}).
\newblock \urlprefix\url{https://doi.org/10.1080/00268976.2017.1333644}.

\bibitem{MISRA2012199}
\bibinfo{author}{Misra, P.~K.}
\newblock \bibinfo{title}{Chapter 7 - Electron–Electron Interaction}.
\newblock In \bibinfo{editor}{Misra, P.~K.} (ed.)
  \emph{\bibinfo{booktitle}{Physics of Condensed Matter}},
  \bibinfo{pages}{199--242} (\bibinfo{publisher}{Academic Press},
  \bibinfo{address}{Boston}, \bibinfo{year}{2012}).
\newblock
  \urlprefix\url{http://www.sciencedirect.com/science/article/pii/B9780123849540000074}.

\bibitem{2005-Heyd-EnergyBand}
\bibinfo{author}{Heyd, J.}, \bibinfo{author}{Peralta, J.~E.},
  \bibinfo{author}{Scuseria, G.~E.} \& \bibinfo{author}{Martin, R.~L.}
\newblock \bibinfo{title}{Energy band gaps and lattice parameters evaluated
  with the Heyd-Scuseria-Ernzerhof screened hybrid functional}.
\newblock \emph{\bibinfo{journal}{J. Chem. Phys.}}
  \textbf{\bibinfo{volume}{123}}, \bibinfo{pages}{174101}
  (\bibinfo{year}{2005}).
\newblock \urlprefix\url{https://doi.org/10.1063/1.2085170}.

\bibitem{2017-Verma-HLE16}
\bibinfo{author}{Verma, P.} \& \bibinfo{author}{Truhlar, D.~G.}
\newblock \bibinfo{title}{HLE16: A Local Kohn–Sham Gradient Approximation
  with Good Performance for Semiconductor Band Gaps and Molecular Excitation
  Energies}.
\newblock \emph{\bibinfo{journal}{J. Phys. Chem. Lett.}}
  \textbf{\bibinfo{volume}{8}}, \bibinfo{pages}{380--387}
  (\bibinfo{year}{2017}).
\newblock \urlprefix\url{https://doi.org/10.1021/acs.jpclett.6b02757}.

\bibitem{arbuznikov_hybrid_2007}
\bibinfo{author}{Arbuznikov, A.~V.}
\newblock \bibinfo{title}{Hybrid exchange correlation functionals and
  potentials: Concept elaboration}.
\newblock \emph{\bibinfo{journal}{J. Struct. Chem.}}
  \textbf{\bibinfo{volume}{48}}, \bibinfo{pages}{S1--S31}
  (\bibinfo{year}{2007}).
\newblock \urlprefix\url{https://doi.org/10.1007/s10947-007-0147-0}.

\bibitem{perdew_hybrid}
\bibinfo{author}{Perdew, J.~P.}, \bibinfo{author}{Ernzerhof, M.} \&
  \bibinfo{author}{Burke, K.}
\newblock \bibinfo{title}{Rationale for mixing exact exchange with density
  functional approximations}.
\newblock \emph{\bibinfo{journal}{J. Chem. Phys.}}
  \textbf{\bibinfo{volume}{105}}, \bibinfo{pages}{9982--9985}
  (\bibinfo{year}{1996}).
\newblock \urlprefix\url{https://doi.org/10.1063/1.472933}.

\bibitem{becke_hybrid}
\bibinfo{author}{Becke, A.~D.}
\newblock \bibinfo{title}{A new mixing of Hartree–Fock and local
  density‐functional theories}.
\newblock \emph{\bibinfo{journal}{J. Chem. Phys.}}
  \textbf{\bibinfo{volume}{98}}, \bibinfo{pages}{1372--1377}
  (\bibinfo{year}{1993}).
\newblock \urlprefix\url{https://doi.org/10.1063/1.464304}.

\bibitem{heyd_hybrid}
\bibinfo{author}{Heyd, J.}, \bibinfo{author}{Scuseria, G.~E.} \&
  \bibinfo{author}{Ernzerhof, M.}
\newblock \bibinfo{title}{Hybrid functionals based on a screened Coulomb
  potential}.
\newblock \emph{\bibinfo{journal}{J. Chem. Phys.}}
  \textbf{\bibinfo{volume}{118}}, \bibinfo{pages}{8207--8215}
  (\bibinfo{year}{2003}).
\newblock \urlprefix\url{https://doi.org/10.1063/1.1564060}.

\bibitem{DMFT-georges-kotliar1}
\bibinfo{author}{Georges, A.}, \bibinfo{author}{Kotliar, G.},
  \bibinfo{author}{Krauth, W.} \& \bibinfo{author}{Rozenberg, M.~J.}
\newblock \bibinfo{title}{Dynamical mean-field theory of strongly correlated
  fermion systems and the limit of infinite dimensions}.
\newblock \emph{\bibinfo{journal}{Rev. Mod. Phys.}}
  \textbf{\bibinfo{volume}{68}}, \bibinfo{pages}{13--125}
  (\bibinfo{year}{1996}).
\newblock \urlprefix\url{https://link.aps.org/doi/10.1103/RevModPhys.68.13}.

\bibitem{DMFT-georges-kotliar-prb}
\bibinfo{author}{Georges, A.} \& \bibinfo{author}{Kotliar, G.}
\newblock \bibinfo{title}{Hubbard model in infinite dimensions}.
\newblock \emph{\bibinfo{journal}{Phys. Rev. B}} \textbf{\bibinfo{volume}{45}},
  \bibinfo{pages}{6479--6483} (\bibinfo{year}{1992}).
\newblock \urlprefix\url{https://link.aps.org/doi/10.1103/PhysRevB.45.6479}.

\bibitem{kotliar2004strongly}
\bibinfo{author}{Kotliar, G.} \& \bibinfo{author}{Vollhardt, D.}
\newblock \bibinfo{title}{Strongly correlated materials: Insights from
  dynamical mean-field theory}.
\newblock \emph{\bibinfo{journal}{Phys. Today}} \textbf{\bibinfo{volume}{57}},
  \bibinfo{pages}{53--60} (\bibinfo{year}{2004}).
\newblock \urlprefix\url{https://doi.org/10.1063/1.1712502}.

\bibitem{kotliar2006electronic}
\bibinfo{author}{Kotliar, G.} \emph{et~al.}
\newblock \bibinfo{title}{Electronic structure calculations with dynamical
  mean-field theory}.
\newblock \emph{\bibinfo{journal}{Rev. Mod. Phys.}}
  \textbf{\bibinfo{volume}{78}}, \bibinfo{pages}{865--951}
  (\bibinfo{year}{2006}).
\newblock \urlprefix\url{https://link.aps.org/doi/10.1103/RevModPhys.78.865}.

\bibitem{georges2004strongly}
\bibinfo{author}{Georges, A.}
\newblock \bibinfo{title}{Strongly Correlated Electron Materials: Dynamical
  Mean‐Field Theory and Electronic Structure}.
\newblock \emph{\bibinfo{journal}{AIP Conf. Proc.}}
  \textbf{\bibinfo{volume}{715}}, \bibinfo{pages}{3--74}
  (\bibinfo{year}{2004}).
\newblock \urlprefix\url{https://aip.scitation.org/doi/abs/10.1063/1.1800733}.

\bibitem{Vollhardt_2011}
\bibinfo{author}{Vollhardt, D.}, \bibinfo{author}{Byczuk, K.} \&
  \bibinfo{author}{Kollar, M.}
\newblock \emph{\bibinfo{title}{Dynamical Mean-Field Theory}},
  \bibinfo{pages}{203--236} (\bibinfo{publisher}{Springer Berlin Heidelberg},
  \bibinfo{address}{Berlin, Heidelberg}, \bibinfo{year}{2012}).
\newblock \urlprefix\url{https://doi.org/10.1007/978-3-642-21831-6_7}.

\bibitem{DFTwDMFT_uthpala}
\bibinfo{title}{DMFTwDFT: An open-source code combining Dynamical Mean Field
  Theory with various density functional theory packages}.
\newblock \emph{\bibinfo{journal}{Comput. Phys. Commun.}}
  \textbf{\bibinfo{volume}{261}}, \bibinfo{pages}{107778}
  (\bibinfo{year}{2021}).
\newblock
  \urlprefix\url{http://www.sciencedirect.com/science/article/pii/S001046552030388X}.

\bibitem{dft_dmft1}
\bibinfo{author}{Paul, A.} \& \bibinfo{author}{Birol, T.}
\newblock \bibinfo{title}{Applications of DFT + DMFT in Materials Science}.
\newblock \emph{\bibinfo{journal}{Annu. Rev. Mater. Res.}}
  \textbf{\bibinfo{volume}{49}}, \bibinfo{pages}{31--52}
  (\bibinfo{year}{2019}).
\newblock \urlprefix\url{https://doi.org/10.1146/annurev-matsci-070218-121825}.

\bibitem{kent2018dft_dmft}
\bibinfo{author}{Kent, P.~R.} \& \bibinfo{author}{Kotliar, G.}
\newblock \bibinfo{title}{Toward a predictive theory of correlated materials}.
\newblock \emph{\bibinfo{journal}{Science}} \textbf{\bibinfo{volume}{361}},
  \bibinfo{pages}{348--354} (\bibinfo{year}{2018}).
\newblock
  \urlprefix\url{https://www.science.org/doi/abs/10.1126/science.aat5975}.

\bibitem{haule2015dft_dmft1}
\bibinfo{author}{Haule, K.} \& \bibinfo{author}{Birol, T.}
\newblock \bibinfo{title}{Free energy from stationary implementation of the
  DFT+ DMFT functional}.
\newblock \emph{\bibinfo{journal}{Phys. Rev. Lett.}}
  \textbf{\bibinfo{volume}{115}}, \bibinfo{pages}{256402}
  (\bibinfo{year}{2015}).
\newblock
  \urlprefix\url{https://link.aps.org/doi/10.1103/PhysRevLett.115.256402}.

\bibitem{haule2015dft_dmft2}
\bibinfo{author}{Haule, K.}
\newblock \bibinfo{title}{Exact double counting in combining the dynamical mean
  field theory and the density functional theory}.
\newblock \emph{\bibinfo{journal}{Phys. Rev. Lett.}}
  \textbf{\bibinfo{volume}{115}}, \bibinfo{pages}{196403}
  (\bibinfo{year}{2015}).
\newblock
  \urlprefix\url{https://link.aps.org/doi/10.1103/PhysRevLett.115.196403}.

\bibitem{koccer2020dft_dmft}
\bibinfo{author}{Ko\ifmmode~\mbox{\c{c}}\else \c{c}\fi{}er, C.~P.},
  \bibinfo{author}{Haule, K.}, \bibinfo{author}{Pascut, G.~L.} \&
  \bibinfo{author}{Monserrat, B.}
\newblock \bibinfo{title}{Efficient lattice dynamics calculations for
  correlated materials with DFT+ DMFT}.
\newblock \emph{\bibinfo{journal}{Phys. Rev. B}}
  \textbf{\bibinfo{volume}{102}}, \bibinfo{pages}{245104}
  (\bibinfo{year}{2020}).
\newblock \urlprefix\url{https://link.aps.org/doi/10.1103/PhysRevB.102.245104}.

\bibitem{aichhorn2016dft_dmft}
\bibinfo{author}{Aichhorn, M.} \emph{et~al.}
\newblock \bibinfo{title}{TRIQS/DFTTools: A TRIQS application for ab initio
  calculations of correlated materials}.
\newblock \emph{\bibinfo{journal}{Comput. Phys. Commun.}}
  \textbf{\bibinfo{volume}{204}}, \bibinfo{pages}{200--208}
  (\bibinfo{year}{2016}).
\newblock
  \urlprefix\url{https://www.sciencedirect.com/science/article/pii/S0010465516300728}.

\bibitem{vollhardt2019dft_dmft}
\bibinfo{author}{Vollhardt, D.}, \bibinfo{author}{Anisimov, V.},
  \bibinfo{author}{Skornyakov, S.} \& \bibinfo{author}{Leonov, I.}
\newblock \bibinfo{title}{Dynamical mean-field theory for correlated electron
  materials}.
\newblock \emph{\bibinfo{journal}{Mater. Today: Proc.}}
  \textbf{\bibinfo{volume}{14}}, \bibinfo{pages}{176--180}
  (\bibinfo{year}{2019}).
\newblock
  \urlprefix\url{https://www.sciencedirect.com/science/article/pii/S2214785319309022}.

\bibitem{DFT+U_LDA+U}
\bibinfo{author}{Himmetoglu, B.}, \bibinfo{author}{Floris, A.},
  \bibinfo{author}{de~Gironcoli, S.} \& \bibinfo{author}{Cococcioni, M.}
\newblock \bibinfo{title}{Hubbard-corrected DFT energy functionals: The LDA+$U$
  description of correlated systems}.
\newblock \emph{\bibinfo{journal}{Int. J. Quantum Chem.}}
  \textbf{\bibinfo{volume}{114}}, \bibinfo{pages}{14--49}
  (\bibinfo{year}{2014}).
\newblock
  \urlprefix\url{https://onlinelibrary.wiley.com/doi/abs/10.1002/qua.24521}.

\bibitem{DFT+U}
\bibinfo{author}{Anisimov, V.~I.}, \bibinfo{author}{Aryasetiawan, F.} \&
  \bibinfo{author}{Lichtenstein, A.~I.}
\newblock \bibinfo{title}{First-principles calculations of the electronic
  structure and spectra of strongly correlated systems: the LDA+$U$ method}.
\newblock \emph{\bibinfo{journal}{J. Condens. Matter Phys.}}
  \textbf{\bibinfo{volume}{9}}, \bibinfo{pages}{767--808}
  (\bibinfo{year}{1997}).
\newblock \urlprefix\url{https://doi.org/10.1088/0953-8984/9/4/002}.

\bibitem{1963-Hubbard}
\bibinfo{author}{Hubbard, J.} \& \bibinfo{author}{Flowers, B.~H.}
\newblock \bibinfo{title}{Electron correlations in narrow energy bands}.
\newblock \emph{\bibinfo{journal}{Proc. R. Soc. A}}
  \textbf{\bibinfo{volume}{276}}, \bibinfo{pages}{238--257}
  (\bibinfo{year}{1963}).
\newblock
  \urlprefix\url{https://royalsocietypublishing.org/doi/abs/10.1098/rspa.1963.0204}.

\bibitem{1964-Hubbard2}
\bibinfo{author}{Hubbard, J.} \& \bibinfo{author}{Flowers, B.~H.}
\newblock \bibinfo{title}{Electron Correlations in Narrow Energy Bands. III. An
  Improved Solution}.
\newblock \emph{\bibinfo{journal}{Proc. R. Soc. A}}
  \textbf{\bibinfo{volume}{281}}, \bibinfo{pages}{401--419}
  (\bibinfo{year}{1964}).
\newblock
  \urlprefix\url{https://royalsocietypublishing.org/doi/abs/10.1098/rspa.1964.0190}.

\bibitem{anisimov_dftu1}
\bibinfo{editor}{Anisimov, V.~I.} (ed.) \emph{\bibinfo{title}{Strong Coulomb
  Correlations in Electronic Structure Calculations}} (\bibinfo{publisher}{CRC
  Press}, \bibinfo{address}{London}, \bibinfo{year}{2000}).
\newblock \urlprefix\url{https://doi.org/10.1201/9781482296877}.

\bibitem{anisimov1997first_dftu2}
\bibinfo{author}{Anisimov, V.~I.}, \bibinfo{author}{Aryasetiawan, F.} \&
  \bibinfo{author}{Lichtenstein, A.}
\newblock \bibinfo{title}{First-principles calculations of the electronic
  structure and spectra of strongly correlated systems: the LDA+$U$ method}.
\newblock \emph{\bibinfo{journal}{J. Condens. Matter Phys.}}
  \textbf{\bibinfo{volume}{9}}, \bibinfo{pages}{767--808}
  (\bibinfo{year}{1997}).
\newblock \urlprefix\url{https://doi.org/10.1088/0953-8984/9/4/002}.

\bibitem{wang2006oxidation_fitU}
\bibinfo{author}{Wang, L.}, \bibinfo{author}{Maxisch, T.} \&
  \bibinfo{author}{Ceder, G.}
\newblock \bibinfo{title}{Oxidation energies of transition metal oxides within
  the GGA+$U$ framework}.
\newblock \emph{\bibinfo{journal}{Phys. Rev. B}} \textbf{\bibinfo{volume}{73}},
  \bibinfo{pages}{195107} (\bibinfo{year}{2006}).
\newblock \urlprefix\url{https://link.aps.org/doi/10.1103/PhysRevB.73.195107}.

\bibitem{pickett1998reformulation}
\bibinfo{author}{Pickett, W.~E.}, \bibinfo{author}{Erwin, S.~C.} \&
  \bibinfo{author}{Ethridge, E.~C.}
\newblock \bibinfo{title}{Reformulation of the LDA+ $U$ method for a
  local-orbital basis}.
\newblock \emph{\bibinfo{journal}{Phys. Rev. B}} \textbf{\bibinfo{volume}{58}},
  \bibinfo{pages}{1201--1209} (\bibinfo{year}{1998}).
\newblock \urlprefix\url{https://link.aps.org/doi/10.1103/PhysRevB.58.1201}.

\bibitem{Cococcioni_PhysRevB.71.035105}
\bibinfo{author}{Cococcioni, M.} \& \bibinfo{author}{de~Gironcoli, S.}
\newblock \bibinfo{title}{Linear response approach to the calculation of the
  effective interaction parameters in the LDA+$U$ method}.
\newblock \emph{\bibinfo{journal}{Phys. Rev. B}} \textbf{\bibinfo{volume}{71}},
  \bibinfo{pages}{035105} (\bibinfo{year}{2005}).
\newblock \urlprefix\url{https://link.aps.org/doi/10.1103/PhysRevB.71.035105}.

\bibitem{constrained_RPA_PhysRevB.74.125106}
\bibinfo{author}{Aryasetiawan, F.}, \bibinfo{author}{Karlsson, K.},
  \bibinfo{author}{Jepsen, O.} \& \bibinfo{author}{Sch\"onberger, U.}
\newblock \bibinfo{title}{Calculations of Hubbard $U$ from first-principles}.
\newblock \emph{\bibinfo{journal}{Phys. Rev. B}} \textbf{\bibinfo{volume}{74}},
  \bibinfo{pages}{125106} (\bibinfo{year}{2006}).
\newblock \urlprefix\url{https://link.aps.org/doi/10.1103/PhysRevB.74.125106}.

\bibitem{timrov2018hubbard}
\bibinfo{author}{Timrov, I.}, \bibinfo{author}{Marzari, N.} \&
  \bibinfo{author}{Cococcioni, M.}
\newblock \bibinfo{title}{Hubbard parameters from density-functional
  perturbation theory}.
\newblock \emph{\bibinfo{journal}{Phys. Rev. B}} \textbf{\bibinfo{volume}{98}},
  \bibinfo{pages}{085127} (\bibinfo{year}{2018}).
\newblock \urlprefix\url{https://link.aps.org/doi/10.1103/PhysRevB.98.085127}.

\bibitem{timrov2021self}
\bibinfo{author}{Timrov, I.}, \bibinfo{author}{Marzari, N.} \&
  \bibinfo{author}{Cococcioni, M.}
\newblock \bibinfo{title}{Self-consistent Hubbard parameters from
  density-functional perturbation theory in the ultrasoft and
  projector-augmented wave formulations}.
\newblock \emph{\bibinfo{journal}{Phys. Rev. B}}
  \textbf{\bibinfo{volume}{103}}, \bibinfo{pages}{045141}
  (\bibinfo{year}{2021}).
\newblock \urlprefix\url{https://link.aps.org/doi/10.1103/PhysRevB.103.045141}.

\bibitem{csacsiouglu2011effective}
\bibinfo{author}{\ifmmode \mbox{\c{S}}\else \c{S}\fi{}a\ifmmode
  \mbox{\c{s}}\else \c{s}\fi{}\ifmmode \imath \else \i
  \fi{}o\ifmmode~\breve{g}\else \u{g}\fi{}lu, E.}, \bibinfo{author}{Friedrich,
  C.} \& \bibinfo{author}{Bl\"ugel, S.}
\newblock \bibinfo{title}{Effective Coulomb interaction in transition metals
  from constrained random-phase approximation}.
\newblock \emph{\bibinfo{journal}{Phys. Rev. B}} \textbf{\bibinfo{volume}{83}},
  \bibinfo{pages}{121101} (\bibinfo{year}{2011}).
\newblock \urlprefix\url{https://link.aps.org/doi/10.1103/PhysRevB.83.121101}.

\bibitem{vaugier2012hubbard}
\bibinfo{author}{Vaugier, L.}, \bibinfo{author}{Jiang, H.} \&
  \bibinfo{author}{Biermann, S.}
\newblock \bibinfo{title}{Hubbard $U$ and Hund exchange $J$ in transition metal
  oxides: Screening versus localization trends from constrained random phase
  approximation}.
\newblock \emph{\bibinfo{journal}{Phys. Rev. B}} \textbf{\bibinfo{volume}{86}},
  \bibinfo{pages}{165105} (\bibinfo{year}{2012}).
\newblock \urlprefix\url{https://link.aps.org/doi/10.1103/PhysRevB.86.165105}.

\bibitem{nakamura2021respack}
\bibinfo{author}{Nakamura, K.} \emph{et~al.}
\newblock \bibinfo{title}{RESPACK: An ab initio tool for derivation of
  effective low-energy model of material}.
\newblock \emph{\bibinfo{journal}{Comput. Phys. Commun.}}
  \textbf{\bibinfo{volume}{261}}, \bibinfo{pages}{107781}
  (\bibinfo{year}{2021}).
\newblock
  \urlprefix\url{https://www.sciencedirect.com/science/article/pii/S001046552030391X}.

\bibitem{mosey2007ab}
\bibinfo{author}{Mosey, N.~J.} \& \bibinfo{author}{Carter, E.~A.}
\newblock \bibinfo{title}{Ab initio evaluation of Coulomb and exchange
  parameters for DFT+$U$ calculations}.
\newblock \emph{\bibinfo{journal}{Phys. Rev. B}} \textbf{\bibinfo{volume}{76}},
  \bibinfo{pages}{155123} (\bibinfo{year}{2007}).
\newblock \urlprefix\url{https://link.aps.org/doi/10.1103/PhysRevB.76.155123}.

\bibitem{agapito2015reformulation}
\bibinfo{author}{Agapito, L.~A.}, \bibinfo{author}{Curtarolo, S.} \&
  \bibinfo{author}{Buongiorno~Nardelli, M.}
\newblock \bibinfo{title}{Reformulation of DFT+$U$ as a pseudohybrid hubbard
  density functional for accelerated materials discovery}.
\newblock \emph{\bibinfo{journal}{Phys. Rev. X}} \textbf{\bibinfo{volume}{5}},
  \bibinfo{pages}{011006} (\bibinfo{year}{2015}).
\newblock \urlprefix\url{https://link.aps.org/doi/10.1103/PhysRevX.5.011006}.

\bibitem{QE_linear_response1}
\bibinfo{author}{Giannozzi, P.} \emph{et~al.}
\newblock \bibinfo{title}{Advanced capabilities for materials modelling with
  Quantum ESPRESSO}.
\newblock \emph{\bibinfo{journal}{J. Condens. Matter Phys.}}
  \textbf{\bibinfo{volume}{29}}, \bibinfo{pages}{465901}
  (\bibinfo{year}{2017}).
\newblock \urlprefix\url{https://doi.org/10.1088/1361-648x/aa8f79}.

\bibitem{QE_linear_response2}
\bibinfo{author}{Giannozzi, P.} \emph{et~al.}
\newblock \bibinfo{title}{QUANTUM ESPRESSO: a modular and open-source software
  project for quantum simulations of materials}.
\newblock \emph{\bibinfo{journal}{J. Condens. Matter Phys.}}
  \textbf{\bibinfo{volume}{21}}, \bibinfo{pages}{395502}
  (\bibinfo{year}{2009}).
\newblock \urlprefix\url{https://doi.org/10.1088/0953-8984/21/39/395502}.

\bibitem{segall2002CASTEP}
\bibinfo{author}{Segall, M.} \emph{et~al.}
\newblock \bibinfo{title}{First-principles simulation: ideas, illustrations and
  the CASTEP code}.
\newblock \emph{\bibinfo{journal}{J. Condens. Matter Phys.}}
  \textbf{\bibinfo{volume}{14}}, \bibinfo{pages}{2717--2744}
  (\bibinfo{year}{2002}).
\newblock \urlprefix\url{https://doi.org/10.1088/0953-8984/14/11/301}.

\bibitem{MeredigPRB2010}
\bibinfo{author}{Meredig, B.}, \bibinfo{author}{Thompson, A.},
  \bibinfo{author}{Hansen, H.~A.}, \bibinfo{author}{Wolverton, C.} \&
  \bibinfo{author}{van~de Walle, A.}
\newblock \bibinfo{title}{Method for locating low-energy solutions within
  DFT+$U$}.
\newblock \emph{\bibinfo{journal}{Phys. Rev. B}} \textbf{\bibinfo{volume}{82}},
  \bibinfo{pages}{195128} (\bibinfo{year}{2010}).
\newblock \urlprefix\url{https://link.aps.org/doi/10.1103/PhysRevB.82.195128}.

\bibitem{AllenWatson2014}
\bibinfo{author}{Allen, J.~P.} \& \bibinfo{author}{Watson, G.~W.}
\newblock \bibinfo{title}{Occupation matrix control of d- and f-electron
  localisations using DFT+$U$}.
\newblock \emph{\bibinfo{journal}{Phys. Chem. Chem. Phys.}}
  \textbf{\bibinfo{volume}{16}}, \bibinfo{pages}{21016--21031}
  (\bibinfo{year}{2014}).
\newblock \urlprefix\url{http://dx.doi.org/10.1039/C4CP01083C}.

\bibitem{payne2019PCCP}
\bibinfo{author}{Payne, A.}, \bibinfo{author}{Aveda{\~n}o-Franco, G.},
  \bibinfo{author}{He, X.}, \bibinfo{author}{Bousquet, E.} \&
  \bibinfo{author}{Romero, A.~H.}
\newblock \bibinfo{title}{Optimizing the orbital occupation in the multiple
  minima problem of magnetic materials from the metaheuristic firefly
  algorithm}.
\newblock \emph{\bibinfo{journal}{Phys. Chem. Chem. Phys.}}
  \textbf{\bibinfo{volume}{21}}, \bibinfo{pages}{21932--21941}.
\newblock \urlprefix\url{http://dx.doi.org/10.1039/C9CP03618K}.

\bibitem{kulik2010_bonding_dft}
\bibinfo{author}{Kulik, H.~J.} \& \bibinfo{author}{Marzari, N.}
\newblock \bibinfo{title}{Systematic study of first-row transition-metal
  diatomic molecules: A self-consistent DFT+$U$ approach}.
\newblock \emph{\bibinfo{journal}{J. Chem. Phys.}}
  \textbf{\bibinfo{volume}{133}}, \bibinfo{pages}{114103}
  (\bibinfo{year}{2010}).
\newblock \urlprefix\url{https://doi.org/10.1063/1.3489110}.

\bibitem{kulik2015_bonding_dftu}
\bibinfo{author}{Kulik, H.~J.}
\newblock \bibinfo{title}{Perspective: Treating electron over-delocalization
  with the DFT+$U$ method}.
\newblock \emph{\bibinfo{journal}{J. Chem. Phys.}}
  \textbf{\bibinfo{volume}{142}}, \bibinfo{pages}{240901}
  (\bibinfo{year}{2015}).
\newblock \urlprefix\url{https://doi.org/10.1063/1.4922693}.

\bibitem{lany_zunger_2008_bonding_dft}
\bibinfo{author}{Lany, S.} \& \bibinfo{author}{Zunger, A.}
\newblock \bibinfo{title}{Assessment of correction methods for the band-gap
  problem and for finite-size effects in supercell defect calculations: Case
  studies for \ce{ZnO} and \ce{GaAs}}.
\newblock \emph{\bibinfo{journal}{Phys. Rev. B}} \textbf{\bibinfo{volume}{78}},
  \bibinfo{pages}{235104} (\bibinfo{year}{2008}).
\newblock \urlprefix\url{https://link.aps.org/doi/10.1103/PhysRevB.78.235104}.

\bibitem{1990-Gelfand-Sampling}
\bibinfo{author}{Gelfand, A.~E.} \& \bibinfo{author}{Smith, A. F.~M.}
\newblock \bibinfo{title}{Sampling-Based Approaches to Calculating Marginal
  Densities}.
\newblock \emph{\bibinfo{journal}{J. Am. Stat. Assoc.}}
  \textbf{\bibinfo{volume}{85}}, \bibinfo{pages}{398--409}
  (\bibinfo{year}{1990}).
\newblock \urlprefix\url{http://www.jstor.org/stable/2289776}.

\bibitem{2006-Jones-fixed_MCMC}
\bibinfo{author}{Jones, G.~L.}, \bibinfo{author}{Haran, M.},
  \bibinfo{author}{Caffo, B.~S.} \& \bibinfo{author}{Neath, R.}
\newblock \bibinfo{title}{Fixed-Width Output Analysis for Markov Chain Monte
  Carlo}.
\newblock \emph{\bibinfo{journal}{J. Am. Stat. Assoc.}}
  \textbf{\bibinfo{volume}{101}}, \bibinfo{pages}{1537--1547}
  (\bibinfo{year}{2006}).
\newblock \urlprefix\url{http://www.jstor.org/stable/27639771}.

\bibitem{mebane2013PCCP}
\bibinfo{author}{Mebane, D.~S.} \emph{et~al.}
\newblock \bibinfo{title}{Bayesian calibration of thermodynamic models for the
  uptake of \ce{CO2} in supported amine sorbents using \textit{ab initio}
  priors}.
\newblock \emph{\bibinfo{journal}{Phys. Chem. Chem. Phys.}}
  \textbf{\bibinfo{volume}{15}}, \bibinfo{pages}{4355--4366}
  (\bibinfo{year}{2013}).
\newblock \urlprefix\url{http://dx.doi.org/10.1039/C3CP42963F}.

\bibitem{Lib_xc_Marques_2012}
\bibinfo{author}{Marques, M.~A.}, \bibinfo{author}{Oliveira, M.~J.} \&
  \bibinfo{author}{Burnus, T.}
\newblock \bibinfo{title}{Libxc: A library of exchange and correlation
  functionals for density functional theory}.
\newblock \emph{\bibinfo{journal}{Comput. Phys. Commun.}}
  \textbf{\bibinfo{volume}{183}}, \bibinfo{pages}{2272--2281}
  (\bibinfo{year}{2012}).
\newblock
  \urlprefix\url{http://www.sciencedirect.com/science/article/pii/S0010465512001750}.

\bibitem{libxc_recent_LEHTOLA20181}
\bibinfo{title}{Recent developments in libxc — A comprehensive library of
  functionals for density functional theory}.
\newblock \emph{\bibinfo{journal}{SoftwareX}} \textbf{\bibinfo{volume}{7}},
  \bibinfo{pages}{1--5} (\bibinfo{year}{2018}).
\newblock
  \urlprefix\url{http://www.sciencedirect.com/science/article/pii/S2352711017300602}.

\bibitem{LDA}
\bibinfo{author}{Ceperley, D.~M.} \& \bibinfo{author}{Alder, B.~J.}
\newblock \bibinfo{title}{Ground State of the Electron Gas by a Stochastic
  Method}.
\newblock \emph{\bibinfo{journal}{Phys. Rev. Lett.}}
  \textbf{\bibinfo{volume}{45}}, \bibinfo{pages}{566--569}
  (\bibinfo{year}{1980}).
\newblock \urlprefix\url{https://link.aps.org/doi/10.1103/PhysRevLett.45.566}.

\bibitem{LDA-jones1989density}
\bibinfo{author}{Jones, R.~O.} \& \bibinfo{author}{Gunnarsson, O.}
\newblock \bibinfo{title}{The density functional formalism, its applications
  and prospects}.
\newblock \emph{\bibinfo{journal}{Rev. Mod. Phys.}}
  \textbf{\bibinfo{volume}{61}}, \bibinfo{pages}{689--746}
  (\bibinfo{year}{1989}).
\newblock \urlprefix\url{https://link.aps.org/doi/10.1103/RevModPhys.61.689}.

\bibitem{LDA_ceperley1980ground}
\bibinfo{author}{Ceperley, D.~M.} \& \bibinfo{author}{Alder, B.~J.}
\newblock \bibinfo{title}{Ground State of the Electron Gas by a Stochastic
  Method}.
\newblock \emph{\bibinfo{journal}{Phys. Rev. Lett.}}
  \textbf{\bibinfo{volume}{45}}, \bibinfo{pages}{566--569}
  (\bibinfo{year}{1980}).
\newblock \urlprefix\url{https://link.aps.org/doi/10.1103/PhysRevLett.45.566}.

\bibitem{LSDA_vonBarth}
\bibinfo{author}{von Barth, U.} \& \bibinfo{author}{Hedin, L.}
\newblock \bibinfo{title}{A local exchange-correlation potential for the spin
  polarized case. i}.
\newblock \emph{\bibinfo{journal}{J. phys., C, Solid state phys.}}
  \textbf{\bibinfo{volume}{5}}, \bibinfo{pages}{1629--1642}
  (\bibinfo{year}{1972}).
\newblock \urlprefix\url{https://doi.org/10.1088/0022-3719/5/13/012}.

\bibitem{gupta_principles_2015}
\bibinfo{author}{Gupta, V.~P.}
\newblock \emph{\bibinfo{title}{Chapter 5 - Density Functional Theory (DFT) and
  Time Dependent DFT (TDDFT)}}, \bibinfo{pages}{155--194}
  (\bibinfo{publisher}{Academic Press}, \bibinfo{address}{Boston},
  \bibinfo{year}{2016}).
\newblock
  \urlprefix\url{http://www.sciencedirect.com/science/article/pii/B9780128034781000054}.

\bibitem{PBE-PRL}
\bibinfo{author}{Perdew, J.~P.}, \bibinfo{author}{Burke, K.} \&
  \bibinfo{author}{Ernzerhof, M.}
\newblock \bibinfo{title}{Perdew, Burke, and Ernzerhof Reply:}.
\newblock \emph{\bibinfo{journal}{Phys. Rev. Lett.}}
  \textbf{\bibinfo{volume}{80}}, \bibinfo{pages}{891--891}
  (\bibinfo{year}{1998}).
\newblock \urlprefix\url{https://link.aps.org/doi/10.1103/PhysRevLett.80.891}.

\bibitem{wentzcovitch_theoretical_2018}
\bibinfo{editor}{Wentzcovitch, R.~M.} \& \bibinfo{editor}{Stixrude, L.} (eds.)
  \emph{\bibinfo{title}{Theoretical and Computational Methods in Mineral
  Physics: Geophysical Applications}} (\bibinfo{publisher}{De Gruyter},
  \bibinfo{year}{2018}).
\newblock \urlprefix\url{https://doi.org/10.1515/9781501508448}.

\bibitem{PBEsol_PhysRevLett.100.136406}
\bibinfo{author}{Perdew, J.~P.} \emph{et~al.}
\newblock \bibinfo{title}{Restoring the Density-Gradient Expansion for Exchange
  in Solids and Surfaces}.
\newblock \emph{\bibinfo{journal}{Phys. Rev. Lett.}}
  \textbf{\bibinfo{volume}{100}}, \bibinfo{pages}{136406}
  (\bibinfo{year}{2008}).
\newblock
  \urlprefix\url{https://link.aps.org/doi/10.1103/PhysRevLett.100.136406}.

\bibitem{PBE_PBEsol_structuralcomp_dongho_nguimdo_density_2015}
\bibinfo{author}{Dongho~Nguimdo, G.~M.} \& \bibinfo{author}{Joubert, D.~P.}
\newblock \bibinfo{title}{A density functional (PBE, PBEsol, HSE06) study of
  the structural, electronic and optical properties of the ternary compounds
  \ce{AgAlX2} (X = S, Se, Te)}.
\newblock \emph{\bibinfo{journal}{Eur. Phys. J. B}}
  \textbf{\bibinfo{volume}{88}}, \bibinfo{pages}{113} (\bibinfo{year}{2015}).
\newblock \urlprefix\url{https://doi.org/10.1140/epjb/e2015-50478-x}.

\bibitem{zhang2018performance}
\bibinfo{author}{Zhang, G.-X.}, \bibinfo{author}{Reilly, A.~M.},
  \bibinfo{author}{Tkatchenko, A.} \& \bibinfo{author}{Scheffler, M.}
\newblock \bibinfo{title}{Performance of various density-functional
  approximations for cohesive properties of 64 bulk solids}.
\newblock \emph{\bibinfo{journal}{New J. Phys.}} \textbf{\bibinfo{volume}{20}},
  \bibinfo{pages}{063020} (\bibinfo{year}{2018}).
\newblock \urlprefix\url{https://doi.org/10.1088/1367-2630/aac7f0}.

\bibitem{de2011performance}
\bibinfo{author}{De~La~Pierre, M.} \emph{et~al.}
\newblock \bibinfo{title}{Performance of six functionals (LDA, PBE, PBESOL,
  B3LYP, PBE0, and WC1LYP) in the simulation of vibrational and dielectric
  properties of crystalline compounds. The case of forsterite \ce{Mg2SiO4}}.
\newblock \emph{\bibinfo{journal}{J. Comput. Chem.}}
  \textbf{\bibinfo{volume}{32}}, \bibinfo{pages}{1775--1784}
  (\bibinfo{year}{2011}).
\newblock
  \urlprefix\url{https://onlinelibrary.wiley.com/doi/abs/10.1002/jcc.21750}.

\bibitem{hinuma2017comparison}
\bibinfo{author}{Hinuma, Y.}, \bibinfo{author}{Hayashi, H.},
  \bibinfo{author}{Kumagai, Y.}, \bibinfo{author}{Tanaka, I.} \&
  \bibinfo{author}{Oba, F.}
\newblock \bibinfo{title}{Comparison of approximations in density functional
  theory calculations: Energetics and structure of binary oxides}.
\newblock \emph{\bibinfo{journal}{Phys. Rev. B}} \textbf{\bibinfo{volume}{96}},
  \bibinfo{pages}{094102} (\bibinfo{year}{2017}).
\newblock \urlprefix\url{https://link.aps.org/doi/10.1103/PhysRevB.96.094102}.

\bibitem{liechenstein_PRB}
\bibinfo{author}{Liechtenstein, A.~I.}, \bibinfo{author}{Anisimov, V.~I.} \&
  \bibinfo{author}{Zaanen, J.}
\newblock \bibinfo{title}{Density-functional theory and strong interactions:
  Orbital ordering in Mott-Hubbard insulators}.
\newblock \emph{\bibinfo{journal}{Phys. Rev. B}} \textbf{\bibinfo{volume}{52}},
  \bibinfo{pages}{R5467--R5470} (\bibinfo{year}{1995}).
\newblock \urlprefix\url{https://link.aps.org/doi/10.1103/PhysRevB.52.R5467}.

\bibitem{ryee2018effect}
\bibinfo{author}{Ryee, S.} \& \bibinfo{author}{Han, M.~J.}
\newblock \bibinfo{title}{The effect of double counting, spin density, and Hund
  interaction in the different DFT+$U$ functionals}.
\newblock \emph{\bibinfo{journal}{Sci. Rep.}} \textbf{\bibinfo{volume}{8}},
  \bibinfo{pages}{9559} (\bibinfo{year}{2018}).
\newblock \urlprefix\url{https://doi.org/10.1038/s41598-018-27731-4}.

\bibitem{wehling20145}
\bibinfo{author}{Wehling, T.}
\newblock \bibinfo{title}{5 Projectors, Hubbard U, Charge Self-Consistency, and
  Double-Counting}.
\newblock In \bibinfo{editor}{Pavarini, E.}, \bibinfo{editor}{Koch, E.},
  \bibinfo{editor}{Vollhardt, D.} \& \bibinfo{editor}{Lichtenstein, A.} (eds.)
  \emph{\bibinfo{booktitle}{Dmft at 25: Infinite dimensions: Lecture notes of
  the autumn school on correlated electrons 2014}}, vol.~\bibinfo{volume}{4},
  \bibinfo{pages}{5.1--5.23} (\bibinfo{publisher}{Forschungszentrum J\"ulich},
  \bibinfo{address}{J\"ulich}, \bibinfo{year}{2014}).
\newblock \urlprefix\url{http://hdl.handle.net/2128/7937}.

\bibitem{greenwood2012chemistry}
\bibinfo{title}{25 - Iron, Ruthenium and Osmium}.
\newblock In \bibinfo{editor}{Greenwoon, N.} \& \bibinfo{editor}{Earnshaw, A.}
  (eds.) \emph{\bibinfo{booktitle}{Chemistry of the Elements (Second
  Edition)}}, \bibinfo{pages}{1070--1112}
  (\bibinfo{publisher}{Butterworth-Heinemann}, \bibinfo{address}{Oxford},
  \bibinfo{year}{1997}), \bibinfo{edition}{second edition} edn.
\newblock
  \urlprefix\url{http://www.sciencedirect.com/science/article/pii/B9780750633659500316}.

\bibitem{Severin_1995}
\bibinfo{author}{Severin, L.}, \bibinfo{author}{Haggstrom, L.},
  \bibinfo{author}{Nordstrom, L.}, \bibinfo{author}{Andersson, Y.} \&
  \bibinfo{author}{Johansson, B.}
\newblock \bibinfo{title}{Magnetism and crystal structure in orthorhombic
  \ce{Fe2P}: a theoretical and experimental study}.
\newblock \emph{\bibinfo{journal}{J. Condens. Matter Phys.}}
  \textbf{\bibinfo{volume}{7}}, \bibinfo{pages}{185--198}
  (\bibinfo{year}{1995}).
\newblock \urlprefix\url{https://doi.org/10.1088/0953-8984/7/1/016}.

\bibitem{Drijver_1976}
\bibinfo{author}{Drijver, J.~W.}, \bibinfo{author}{Sinnema, S.~G.} \&
  \bibinfo{author}{van~der Woude, F.}
\newblock \bibinfo{title}{Magnetic properties of hexagonal and cubic
  \ce{Fe3Ge}}.
\newblock \emph{\bibinfo{journal}{J. Phys. F: Met. Phys.}}
  \textbf{\bibinfo{volume}{6}}, \bibinfo{pages}{2165--2177}
  (\bibinfo{year}{1976}).
\newblock \urlprefix\url{https://doi.org/10.1088/0305-4608/6/11/015}.

\bibitem{hayashi_BaFeO3}
\bibinfo{author}{Hayashi, N.} \emph{et~al.}
\newblock \bibinfo{title}{\ce{BaFeO3}: A Ferromagnetic Iron Oxide}.
\newblock \emph{\bibinfo{journal}{Angew. Chem.}}
  \textbf{\bibinfo{volume}{123}}, \bibinfo{pages}{12755--12758}
  (\bibinfo{year}{2011}).
\newblock
  \urlprefix\url{https://onlinelibrary.wiley.com/doi/abs/10.1002/anie.201105276}.

\bibitem{zhao_SrFeO3}
\bibinfo{author}{Zhao, Y.} \& \bibinfo{author}{Zhou, P.}
\newblock \bibinfo{title}{Metal-insulator transition in helical
  \ce{BaFeO_{3$-\delta$}} antiferromagnet}.
\newblock \emph{\bibinfo{journal}{J. Magn. Magn. Mater.}}
  \textbf{\bibinfo{volume}{281}}, \bibinfo{pages}{214--220}
  (\bibinfo{year}{2004}).
\newblock
  \urlprefix\url{https://www.sciencedirect.com/science/article/pii/S0304885304005517}.

\bibitem{mori_BaFeO3}
\bibinfo{author}{Mori, K.} \emph{et~al.}
\newblock \bibinfo{title}{Mixed magnetic phase in 6\textit{H}-type
  \ce{BaFeO_{3$-\delta$}}}.
\newblock \emph{\bibinfo{journal}{J. Appl. Crystallogr.}}
  \textbf{\bibinfo{volume}{40}}, \bibinfo{pages}{s501--s505}
  (\bibinfo{year}{2007}).
\newblock \urlprefix\url{https://doi.org/10.1107/S0021889807001653}.

\bibitem{nortonBaFeO3_report}
\bibinfo{author}{Norton, D.~P.}
\newblock \bibinfo{title}{Synthesis and Characterization of \ce{BaFeO3}, (Ba,
  Bi)\ce{FeO3}, and Related Epitaxial Thin Films and Nanostructures}.
\newblock \bibinfo{type}{Tech. Rep.}, \bibinfo{institution}{Dept. of Materials
  Science and Engr., University of Florida, Gainesville, FL}
  (\bibinfo{year}{2009}).
\newblock \eprint{https://apps.dtic.mil/sti/pdfs/ADA510215.pdf}.

\bibitem{BaFeO3_mag_band_PRB}
\bibinfo{author}{Tsuyama, T.} \emph{et~al.}
\newblock \bibinfo{title}{X-ray spectroscopic study of \ce{BaFeO3} thin films:
  An \ce{Fe4+} ferromagnetic insulator}.
\newblock \emph{\bibinfo{journal}{Phys. Rev. B}} \textbf{\bibinfo{volume}{91}},
  \bibinfo{pages}{115101} (\bibinfo{year}{2015}).
\newblock \urlprefix\url{https://link.aps.org/doi/10.1103/PhysRevB.91.115101}.

\bibitem{ishiwata_SrFeO3}
\bibinfo{author}{Ishiwata, S.} \emph{et~al.}
\newblock \bibinfo{title}{Versatile helimagnetic phases under magnetic fields
  in cubic perovskite \ce{SrFeO3}}.
\newblock \emph{\bibinfo{journal}{Phys. Rev. B}} \textbf{\bibinfo{volume}{84}},
  \bibinfo{pages}{054427} (\bibinfo{year}{2011}).
\newblock \urlprefix\url{https://link.aps.org/doi/10.1103/PhysRevB.84.054427}.

\bibitem{gap_SrFeO3}
\bibinfo{author}{Ghaffari, M.}, \bibinfo{author}{Huang, H.},
  \bibinfo{author}{Tan, O.~K.} \& \bibinfo{author}{Shannon, M.}
\newblock \bibinfo{title}{Band gap measurement of \ce{SrFeO_{3$-\delta$}} by
  ultraviolet photoelectron spectroscopy and photovoltage method}.
\newblock \emph{\bibinfo{journal}{CrystEngComm}} \textbf{\bibinfo{volume}{14}},
  \bibinfo{pages}{7487--7492} (\bibinfo{year}{2012}).
\newblock \urlprefix\url{http://dx.doi.org/10.1039/C2CE25751C}.

\bibitem{bousquet2010j}
\bibinfo{author}{Bousquet, E.} \& \bibinfo{author}{Spaldin, N.}
\newblock \bibinfo{title}{J dependence in the LSDA+$U$ treatment of
  noncollinear magnets}.
\newblock \emph{\bibinfo{journal}{Phys. Rev. B}} \textbf{\bibinfo{volume}{82}},
  \bibinfo{pages}{220402} (\bibinfo{year}{2010}).
\newblock \urlprefix\url{https://link.aps.org/doi/10.1103/PhysRevB.82.220402}.

\bibitem{himmetoglu2014IJQC}
\bibinfo{author}{Himmetoglu, B.}, \bibinfo{author}{Floris, A.},
  \bibinfo{author}{De~Gironcoli, S.} \& \bibinfo{author}{Cococcioni, M.}
\newblock \bibinfo{title}{Hubbard-corrected DFT energy functionals: The LDA+$U$
  description of correlated systems}.
\newblock \emph{\bibinfo{journal}{Int. J. Quantum Chem.}}
  \textbf{\bibinfo{volume}{114}}, \bibinfo{pages}{14--49}
  (\bibinfo{year}{2014}).
\newblock
  \urlprefix\url{https://onlinelibrary.wiley.com/doi/abs/10.1002/qua.24521}.

\bibitem{nakamura2009first}
\bibinfo{author}{Nakamura, H.}, \bibinfo{author}{Hayashi, N.},
  \bibinfo{author}{Nakai, N.}, \bibinfo{author}{Okumura, M.} \&
  \bibinfo{author}{Machida, M.}
\newblock \bibinfo{title}{First-principle electronic structure calculations for
  magnetic moment in iron-based superconductors: An LSDA+negative $U$ study}.
\newblock \emph{\bibinfo{journal}{Physica C Supercond}}
  \textbf{\bibinfo{volume}{469}}, \bibinfo{pages}{908--911}
  (\bibinfo{year}{2009}).
\newblock
  \urlprefix\url{https://www.sciencedirect.com/science/article/pii/S0921453409001804}.

\bibitem{KDE_scott}
\bibinfo{author}{Scott, D.~W.}
\newblock \bibinfo{title}{On optimal and data-based histograms}.
\newblock \emph{\bibinfo{journal}{Biometrika}} \textbf{\bibinfo{volume}{66}},
  \bibinfo{pages}{605--610} (\bibinfo{year}{1979}).
\newblock \urlprefix\url{https://doi.org/10.1093/biomet/66.3.605}.

\bibitem{dudarev}
\bibinfo{author}{Dudarev, S.~L.}, \bibinfo{author}{Botton, G.~A.},
  \bibinfo{author}{Savrasov, S.~Y.}, \bibinfo{author}{Humphreys, C.~J.} \&
  \bibinfo{author}{Sutton, A.~P.}
\newblock \bibinfo{title}{Electron-energy-loss spectra and the structural
  stability of nickel oxide: An LSDA+$U$ study}.
\newblock \emph{\bibinfo{journal}{Phys. Rev. B}} \textbf{\bibinfo{volume}{57}},
  \bibinfo{pages}{1505--1509} (\bibinfo{year}{1998}).
\newblock \urlprefix\url{https://link.aps.org/doi/10.1103/PhysRevB.57.1505}.

\bibitem{Fe_structure_LandoltBornstein1994}
\bibinfo{author}{Chiarotti, G.}
\newblock \bibinfo{title}{1.6 Crystal structures and bulk lattice parameters of
  materials quoted in the volume}.
\newblock In \bibinfo{editor}{Chiarotti, G.} (ed.)
  \emph{\bibinfo{booktitle}{Physics of Solid Surfaces · Structure}},
  \bibinfo{pages}{21--26} (\bibinfo{publisher}{Springer},
  \bibinfo{address}{Berlin Heidelberg}, \bibinfo{year}{1995}).
\newblock
  \urlprefix\url{https://materials.springer.com/lb/docs/sm_lbs_978-3-540-47397-8_6}.

\bibitem{cornell2004FeOxides_ch6}
\bibinfo{author}{Cornell, R.~M.} \& \bibinfo{author}{Schwertmann, U.}
\newblock \emph{\bibinfo{title}{Electronic, Electrical and Magnetic Properties
  and Colour}}, chap.~\bibinfo{chapter}{6}, \bibinfo{pages}{111--137}
  (\bibinfo{publisher}{John Wiley \& Sons, Ltd}, \bibinfo{address}{Weinheim},
  \bibinfo{year}{2004}).
\newblock
  \urlprefix\url{https://onlinelibrary.wiley.com/doi/abs/10.1002/3527602097.ch6}.

\bibitem{Fe_magMom}
\bibinfo{author}{Kikuchi, H.}, \bibinfo{author}{Suzuki, Y.} \&
  \bibinfo{author}{Katayama, T.}
\newblock \bibinfo{title}{Structure and magnetic properties of single‐crystal
  Fe/Au(100) superlattices synthesized using RHEED oscillation}.
\newblock \emph{\bibinfo{journal}{Int. J. Appl. Phys.}}
  \textbf{\bibinfo{volume}{67}}, \bibinfo{pages}{5403--5405}
  (\bibinfo{year}{1990}).
\newblock \urlprefix\url{https://doi.org/10.1063/1.344567}.

\bibitem{TOBOLA1996708}
\bibinfo{author}{Tobola, J.} \emph{et~al.}
\newblock \bibinfo{title}{Magnetism of \ce{Fe2P} investigated by neutron
  experiments and band structure calculations}.
\newblock \emph{\bibinfo{journal}{J. Magn. Magn. Mater.}}
  \textbf{\bibinfo{volume}{157-158}}, \bibinfo{pages}{708--710}
  (\bibinfo{year}{1996}).
\newblock
  \urlprefix\url{http://www.sciencedirect.com/science/article/pii/0304885395012583}.

\bibitem{gap_Fe2P_SUGIZAKI201750}
\bibinfo{author}{Sugizaki, Y.}, \bibinfo{author}{Motoyama, H.},
  \bibinfo{author}{Edamoto, K.} \& \bibinfo{author}{Ozawa, K.}
\newblock \bibinfo{title}{Electronic structure of \ce{Fe$_2$P}(10$\bar{1}$0)
  studied by soft X-ray photoelectron spectroscopy and X-ray absorption
  spectroscopy}.
\newblock \emph{\bibinfo{journal}{Surf. Sci.}} \textbf{\bibinfo{volume}{664}},
  \bibinfo{pages}{50--55} (\bibinfo{year}{2017}).
\newblock
  \urlprefix\url{http://www.sciencedirect.com/science/article/pii/S0039602817302182}.

\bibitem{BaFeO3_lattice_Taib2016StructuralEA}
\bibinfo{author}{Taib, M.}, \bibinfo{author}{Hussin, N.},
  \bibinfo{author}{Samat, M.}, \bibinfo{author}{Hassan, O.} \&
  \bibinfo{author}{Yahya, M.}
\newblock \bibinfo{title}{Structural, Electronic and Optical Properties of
  \ce{BaTiO3} and \ce{BaFeO3} From First Principles LDA+$U$ Study}.
\newblock \emph{\bibinfo{journal}{Int. J. Electroactive Mater}}
  \textbf{\bibinfo{volume}{4}}, \bibinfo{pages}{14--17} (\bibinfo{year}{2016}).

\bibitem{lattice_SrFeO3_osti_1376467}
\bibinfo{author}{Santana, J.~A.}, \bibinfo{author}{Krogel, J.~T.},
  \bibinfo{author}{Kent, P.~R.} \& \bibinfo{author}{Reboredo, F.~A.}
\newblock \bibinfo{title}{Diffusion quantum Monte Carlo calculations of
  \ce{SrFeO3} and \ce{LaFeO3}}.
\newblock \emph{\bibinfo{journal}{J. Chem. Phys.}}
  \textbf{\bibinfo{volume}{147}}, \bibinfo{pages}{034701}
  (\bibinfo{year}{2017}).
\newblock \urlprefix\url{https://doi.org/10.1063/1.4994083}.

\bibitem{Fe5PB2mcguireFe5SiB2}
\bibinfo{author}{McGuire, M.~A.} \& \bibinfo{author}{Parker, D.~S.}
\newblock \bibinfo{title}{Magnetic and structural properties of ferromagnetic
  \ce{Fe5PB2} and \ce{Fe5SiB2} and effects of Co and Mn substitutions}.
\newblock \emph{\bibinfo{journal}{Int. J. Appl. Phys.}}
  \textbf{\bibinfo{volume}{118}}, \bibinfo{pages}{163903}
  (\bibinfo{year}{2015}).
\newblock \urlprefix\url{https://doi.org/10.1063/1.4934496}.

\bibitem{MATAR1995169}
\bibinfo{author}{Matar, S.}, \bibinfo{author}{Mohn, P.} \&
  \bibinfo{author}{Demazeau, G.}
\newblock \bibinfo{title}{The magnetic structure of \ce{SrFeO3} calculated
  within LDA}.
\newblock \emph{\bibinfo{journal}{J. Magn. Magn. Mater.}}
  \textbf{\bibinfo{volume}{140-144}}, \bibinfo{pages}{169--170}
  (\bibinfo{year}{1995}).
\newblock
  \urlprefix\url{http://www.sciencedirect.com/science/article/pii/030488539401129X}.
\newblock \bibinfo{note}{International Conference on Magnetism}.

\bibitem{lu_room_2010}
\bibinfo{author}{Lu, J.} \emph{et~al.}
\newblock \bibinfo{title}{On the room temperature multiferroic \ce{BiFeO3}:
  magnetic,dielectric and thermal properties}.
\newblock \emph{\bibinfo{journal}{Eur. Phys. J. B}}
  \textbf{\bibinfo{volume}{75}}, \bibinfo{pages}{451--460}
  (\bibinfo{year}{2010}).
\newblock \urlprefix\url{https://doi.org/10.1140/epjb/e2010-00170-x}.

\bibitem{2016ApPhA_SrFeO3}
\bibinfo{author}{Radheep, D.~M.}, \bibinfo{author}{Shanmugapriya, K.},
  \bibinfo{author}{Palanivel, B.} \& \bibinfo{author}{Murugan, R.}
\newblock \bibinfo{title}{Magnetic field-induced switching of magnetic ordering
  in \ce{SrFeO_{3$-\delta$}}}.
\newblock \emph{\bibinfo{journal}{Appl. Phys. A}}
  \textbf{\bibinfo{volume}{122}}, \bibinfo{pages}{778} (\bibinfo{year}{2016}).
\newblock \urlprefix\url{https://doi.org/10.1007/s00339-016-0303-5}.

\bibitem{schrettle2012wustite}
\bibinfo{author}{Schrettle, F.} \emph{et~al.}
\newblock \bibinfo{title}{W{\"u}stite: electric, thermodynamic and optical
  properties of \ce{FeO}}.
\newblock \emph{\bibinfo{journal}{Eur. Phys. J. B}}
  \textbf{\bibinfo{volume}{85}}, \bibinfo{pages}{164} (\bibinfo{year}{2012}).
\newblock \urlprefix\url{https://doi.org/10.1140/epjb/e2012-30201-5}.

\bibitem{goodenough_magnetic_1970}
\bibinfo{editor}{Hellwege, K.-H.} \& \bibinfo{editor}{Hellwege, A.~M.} (eds.)
  \emph{\bibinfo{title}{Magnetic and Other Properties of Oxides and Related
  Compounds}} (\bibinfo{publisher}{Springer}, \bibinfo{address}{Berlin
  Heidelberg}, \bibinfo{year}{1970}).
\newblock
  \urlprefix\url{https://materials.springer.com/lb/docs/sm_lbs_978-3-540-36202-9}.

\bibitem{bowen1975wustite}
\bibinfo{author}{Bowen, H.}, \bibinfo{author}{Adler, D.} \&
  \bibinfo{author}{Auker, B.}
\newblock \bibinfo{title}{Electrical and optical properties of \ce{FeO}}.
\newblock \emph{\bibinfo{journal}{J. Solid State Chem.}}
  \textbf{\bibinfo{volume}{12}}, \bibinfo{pages}{355--359}
  (\bibinfo{year}{1975}).
\newblock
  \urlprefix\url{https://www.sciencedirect.com/science/article/pii/0022459675903400}.

\bibitem{cornell2004FeOxides_ch2}
\bibinfo{author}{Cornell, R.~M.} \& \bibinfo{author}{Schwertmann, U.}
\newblock \emph{\bibinfo{title}{Crystal Structure}}, chap.~\bibinfo{chapter}{2}
  (\bibinfo{publisher}{John Wiley \& Sons, Ltd}, \bibinfo{address}{Weinheim},
  \bibinfo{year}{2004}).
\newblock
  \urlprefix\url{https://onlinelibrary.wiley.com/doi/abs/10.1002/3527602097.ch2}.

\bibitem{1980-Finger-Fe2O3}
\bibinfo{author}{Finger, L.~W.} \& \bibinfo{author}{Hazen, R.~M.}
\newblock \bibinfo{title}{Crystal structure and isothermal compression of
  \ce{Fe2O3}, \ce{Cr2O3}, and \ce{V2O3} to 50 Kbars}.
\newblock \emph{\bibinfo{journal}{Int. J. Appl. Phys.}}
  \textbf{\bibinfo{volume}{51}}, \bibinfo{pages}{5362--5367}
  (\bibinfo{year}{1980}).
\newblock \urlprefix\url{https://aip.scitation.org/doi/abs/10.1063/1.327451}.

\bibitem{Coey_1971}
\bibinfo{author}{Coey, J. M.~D.} \& \bibinfo{author}{Sawatzky, G.~A.}
\newblock \bibinfo{title}{A study of hyperfine interactions in the system
  \ce{(Fe$_{1-x}$Rh$_x$)2O3} using the M\"{o}ssbauer effect (Bonding
  parameters)}.
\newblock \emph{\bibinfo{journal}{J. phys., C, Solid state phys.}}
  \textbf{\bibinfo{volume}{4}}, \bibinfo{pages}{2386--2407}
  (\bibinfo{year}{1971}).
\newblock \urlprefix\url{https://doi.org/10.1088/0022-3719/4/15/025}.

\bibitem{coey2013_mag_oxide}
\bibinfo{author}{Coey, J. M.~D.}, \bibinfo{author}{Venkatesan, M.} \&
  \bibinfo{author}{Xu, H.}
\newblock \emph{\bibinfo{title}{Introduction to Magnetic Oxides}},
  chap.~\bibinfo{chapter}{1}, \bibinfo{pages}{1--49} (\bibinfo{publisher}{John
  Wiley \& Sons, Ltd}, \bibinfo{address}{Weinheim}, \bibinfo{year}{2013}).
\newblock
  \urlprefix\url{https://onlinelibrary.wiley.com/doi/abs/10.1002/9783527654864.ch1}.

\bibitem{lamichhane2018AlFe2B2_mag_lat}
\bibinfo{author}{Lamichhane, T.~N.} \emph{et~al.}
\newblock \bibinfo{title}{Magnetic properties of single crystalline itinerant
  ferromagnet \ce{AlFe2B2}}.
\newblock \emph{\bibinfo{journal}{Phys. Rev. Materials}}
  \textbf{\bibinfo{volume}{2}}, \bibinfo{pages}{084408} (\bibinfo{year}{2018}).
\newblock
  \urlprefix\url{https://link.aps.org/doi/10.1103/PhysRevMaterials.2.084408}.

\bibitem{barua2019AlFe2B2_conductor}
\bibinfo{author}{Barua, R.} \emph{et~al.}
\newblock \bibinfo{title}{Enhanced room-temperature magnetocaloric effect and
  tunable magnetic response in Ga-and Ge-substituted \ce{AlFe2B2}}.
\newblock \emph{\bibinfo{journal}{J. Alloys Compd.}}
  \textbf{\bibinfo{volume}{777}}, \bibinfo{pages}{1030--1038}
  (\bibinfo{year}{2019}).
\newblock
  \urlprefix\url{https://www.sciencedirect.com/science/article/pii/S0925838818338805}.

\bibitem{elmassalami2011AlFe2B2_mag}
\bibinfo{author}{ElMassalami, M.}, \bibinfo{author}{Oliveira, D. d.~S.} \&
  \bibinfo{author}{Takeya, H.}
\newblock \bibinfo{title}{On the ferromagnetism of \ce{AlFe2B2}}.
\newblock \emph{\bibinfo{journal}{J. Magn. Magn. Mater.}}
  \textbf{\bibinfo{volume}{323}}, \bibinfo{pages}{2133--2136}
  (\bibinfo{year}{2011}).
\newblock
  \urlprefix\url{https://www.sciencedirect.com/science/article/pii/S0304885311001661}.

\bibitem{ali2017AlFe2B2_mag}
\bibinfo{author}{Ali, T.}, \bibinfo{author}{Khan, M.}, \bibinfo{author}{Ahmed,
  E.} \& \bibinfo{author}{Ali, A.}
\newblock \bibinfo{title}{Phase analysis of \ce{AlFe2B2} by synchrotron X-ray
  diffraction, magnetic and M{\"o}ssbauer studies}.
\newblock \emph{\bibinfo{journal}{Prog. Nat. Sci. Mater. Int}}
  \textbf{\bibinfo{volume}{27}}, \bibinfo{pages}{251--256}
  (\bibinfo{year}{2017}).
\newblock
  \urlprefix\url{https://www.sciencedirect.com/science/article/pii/S1002007116301319}.

\bibitem{meng2016Fe_oxides}
\bibinfo{author}{Meng, Y.} \emph{et~al.}
\newblock \bibinfo{title}{When density functional approximations meet iron
  oxides}.
\newblock \emph{\bibinfo{journal}{J. Chem. Theory Comput.}}
  \textbf{\bibinfo{volume}{12}}, \bibinfo{pages}{5132--5144}
  (\bibinfo{year}{2016}).
\newblock \urlprefix\url{https://doi.org/10.1021/acs.jctc.6b00640}.

\bibitem{FeO_Subhasish}
\bibinfo{author}{Mandal, S.}, \bibinfo{author}{Haule, K.},
  \bibinfo{author}{Rabe, K.~M.} \& \bibinfo{author}{Vanderbilt, D.}
\newblock \bibinfo{title}{Influence of magnetic ordering on the spectral
  properties of binary transition metal oxides}.
\newblock \emph{\bibinfo{journal}{Phys. Rev. B}}
  \textbf{\bibinfo{volume}{100}}, \bibinfo{pages}{245109}
  (\bibinfo{year}{2019}).
\newblock \urlprefix\url{https://link.aps.org/doi/10.1103/PhysRevB.100.245109}.

\bibitem{FeO_Subhasish_nature}
\bibinfo{author}{Mandal, S.}, \bibinfo{author}{Haule, K.},
  \bibinfo{author}{Rabe, K.~M.} \& \bibinfo{author}{Vanderbilt, D.}
\newblock \bibinfo{title}{Systematic beyond-DFT study of binary transition
  metal oxides}.
\newblock \emph{\bibinfo{journal}{Npj Comput. Mater.}}
  \textbf{\bibinfo{volume}{5}}, \bibinfo{pages}{1--8} (\bibinfo{year}{2019}).
\newblock \urlprefix\url{https://doi.org/10.1038/s41524-019-0251-7}.

\bibitem{kulik-marzari2008pseudo}
\bibinfo{author}{Kulik, H.~J.} \& \bibinfo{author}{Marzari, N.}
\newblock \bibinfo{title}{A self-consistent Hubbard U density-functional theory
  approach to the addition-elimination reactions of hydrocarbons on bare
  \ce{FeO+}}.
\newblock \emph{\bibinfo{journal}{J. Chem. Phys.}}
  \textbf{\bibinfo{volume}{129}}, \bibinfo{pages}{134314}
  (\bibinfo{year}{2008}).
\newblock \urlprefix\url{https://doi.org/10.1063/1.2987444}.

\bibitem{vasp1}
\bibinfo{author}{Kresse, G.} \& \bibinfo{author}{Hafner, J.}
\newblock \bibinfo{title}{Ab initio molecular dynamics for liquid metals}.
\newblock \emph{\bibinfo{journal}{Phys. Rev. B}} \textbf{\bibinfo{volume}{47}},
  \bibinfo{pages}{558--561} (\bibinfo{year}{1993}).
\newblock \urlprefix\url{https://link.aps.org/doi/10.1103/PhysRevB.47.558}.

\bibitem{vasp2}
\bibinfo{author}{Kresse, G.} \& \bibinfo{author}{Hafner, J.}
\newblock \bibinfo{title}{Ab initio molecular-dynamics simulation of the
  liquid-metal--amorphous-semiconductor transition in germanium}.
\newblock \emph{\bibinfo{journal}{Phys. Rev. B}} \textbf{\bibinfo{volume}{49}},
  \bibinfo{pages}{14251--14269} (\bibinfo{year}{1994}).
\newblock \urlprefix\url{https://link.aps.org/doi/10.1103/PhysRevB.49.14251}.

\bibitem{vasp3}
\bibinfo{author}{Kresse, G.} \& \bibinfo{author}{Furthm{\"u}ller, J.}
\newblock \bibinfo{title}{Efficiency of ab-initio total energy calculations for
  metals and semiconductors using a plane-wave basis set}.
\newblock \emph{\bibinfo{journal}{Comput. Mater. Sci.}}
  \textbf{\bibinfo{volume}{6}}, \bibinfo{pages}{15--50} (\bibinfo{year}{1996}).
\newblock
  \urlprefix\url{https://www.sciencedirect.com/science/article/pii/0927025696000080}.

\bibitem{vasp4}
\bibinfo{author}{Kresse, G.} \& \bibinfo{author}{Furthm\"uller, J.}
\newblock \bibinfo{title}{Efficient iterative schemes for ab initio
  total-energy calculations using a plane-wave basis set}.
\newblock \emph{\bibinfo{journal}{Phys. Rev. B}} \textbf{\bibinfo{volume}{54}},
  \bibinfo{pages}{11169--11186} (\bibinfo{year}{1996}).
\newblock \urlprefix\url{https://link.aps.org/doi/10.1103/PhysRevB.54.11169}.

\bibitem{blochl_PAW}
\bibinfo{author}{Bl\"ochl, P.~E.}
\newblock \bibinfo{title}{Projector augmented-wave method}.
\newblock \emph{\bibinfo{journal}{Phys. Rev. B}} \textbf{\bibinfo{volume}{50}},
  \bibinfo{pages}{17953--17979} (\bibinfo{year}{1994}).
\newblock \urlprefix\url{https://link.aps.org/doi/10.1103/PhysRevB.50.17953}.

\bibitem{kresse_PAW}
\bibinfo{author}{Kresse, G.} \& \bibinfo{author}{Joubert, D.}
\newblock \bibinfo{title}{From ultrasoft pseudopotentials to the projector
  augmented-wave method}.
\newblock \emph{\bibinfo{journal}{Phys. Rev. B}} \textbf{\bibinfo{volume}{59}},
  \bibinfo{pages}{1758--1775} (\bibinfo{year}{1999}).
\newblock \urlprefix\url{https://link.aps.org/doi/10.1103/PhysRevB.59.1758}.

\bibitem{monkhorst}
\bibinfo{author}{Monkhorst, H.~J.} \& \bibinfo{author}{Pack, J.~D.}
\newblock \bibinfo{title}{Special points for Brillouin-zone integrations}.
\newblock \emph{\bibinfo{journal}{Phys. Rev. B}} \textbf{\bibinfo{volume}{13}},
  \bibinfo{pages}{5188--5192} (\bibinfo{year}{1976}).
\newblock \urlprefix\url{https://link.aps.org/doi/10.1103/PhysRevB.13.5188}.

\bibitem{bengone2000implementation}
\bibinfo{author}{Bengone, O.}, \bibinfo{author}{Alouani, M.},
  \bibinfo{author}{Bl\"ochl, P.} \& \bibinfo{author}{Hugel, J.}
\newblock \bibinfo{title}{Implementation of the projector augmented-wave
  LDA+$U$ method: Application to the electronic structure of \ce{NiO}}.
\newblock \emph{\bibinfo{journal}{Phys. Rev. B}} \textbf{\bibinfo{volume}{62}},
  \bibinfo{pages}{16392--16401} (\bibinfo{year}{2000}).
\newblock \urlprefix\url{https://link.aps.org/doi/10.1103/PhysRevB.62.16392}.

\bibitem{matplotlib}
\bibinfo{author}{Hunter, J.~D.}
\newblock \bibinfo{title}{Matplotlib: A 2D graphics environment}.
\newblock \emph{\bibinfo{journal}{Comput. Sci. Eng.}}
  \textbf{\bibinfo{volume}{9}}, \bibinfo{pages}{90--95} (\bibinfo{year}{2007}).
\newblock \urlprefix\url{https://doi.org/10.1109/MCSE.2007.55}.

\bibitem{pyvista}
\bibinfo{author}{Sullivan, C.~B.} \& \bibinfo{author}{Kaszynski, A.}
\newblock \bibinfo{title}{{PyVista}: 3D plotting and mesh analysis through a
  streamlined interface for the Visualization Toolkit ({VTK})}.
\newblock \emph{\bibinfo{journal}{J. Open Source Softw.}}
  \textbf{\bibinfo{volume}{4}}, \bibinfo{pages}{1450} (\bibinfo{year}{2019}).
\newblock \urlprefix\url{https://doi.org/10.21105/joss.01450}.

\bibitem{numpy}
\bibinfo{author}{Harris, C.~R.} \emph{et~al.}
\newblock \bibinfo{title}{Array programming with NumPy}.
\newblock \emph{\bibinfo{journal}{Nature}} \textbf{\bibinfo{volume}{585}},
  \bibinfo{pages}{357--362} (\bibinfo{year}{2020}).
\newblock \urlprefix\url{https://doi.org/10.1038/s41586-020-2649-2}.

\bibitem{scipy}
\bibinfo{author}{Virtanen, P.} \emph{et~al.}
\newblock \bibinfo{title}{SciPy 1.0: fundamental algorithms for scientific
  computing in Python}.
\newblock \emph{\bibinfo{journal}{Nat. Methods.}}
  \textbf{\bibinfo{volume}{17}}, \bibinfo{pages}{261--272}
  (\bibinfo{year}{2020}).
\newblock \urlprefix\url{https://doi.org/10.1038/s41592-019-0686-2}.

\bibitem{derondeau2016hyperfine}
\bibinfo{author}{Derondeau, G.}, \bibinfo{author}{Min\'ar, J.} \&
  \bibinfo{author}{Ebert, H.}
\newblock \bibinfo{title}{Hyperfine fields in the \ce{BaFe2As2} family and
  their relation to the magnetic moment}.
\newblock \emph{\bibinfo{journal}{Phys. Rev. B}} \textbf{\bibinfo{volume}{94}},
  \bibinfo{pages}{214508} (\bibinfo{year}{2016}).
\newblock \urlprefix\url{https://link.aps.org/doi/10.1103/PhysRevB.94.214508}.

\bibitem{BaFeAs-crystal}
\bibinfo{author}{Rotter, M.} \emph{et~al.}
\newblock \bibinfo{title}{Spin-density-wave anomaly at 140 K in the ternary
  iron arsenide \ce{BaFe2As2}}.
\newblock \emph{\bibinfo{journal}{Phys. Rev. B}} \textbf{\bibinfo{volume}{78}},
  \bibinfo{pages}{020503} (\bibinfo{year}{2008}).
\newblock \urlprefix\url{https://link.aps.org/doi/10.1103/PhysRevB.78.020503}.

\bibitem{BaFeAs-magmom1}
\bibinfo{author}{Huang, Q.} \emph{et~al.}
\newblock \bibinfo{title}{Neutron-diffraction measurements of magnetic order
  and a structural transition in the parent \ce{BaFe2As2} compound of
  \ce{FeAs}-based high-temperature superconductors}.
\newblock \emph{\bibinfo{journal}{Phys. Rev. Lett.}}
  \textbf{\bibinfo{volume}{101}}, \bibinfo{pages}{257003}
  (\bibinfo{year}{2008}).
\newblock
  \urlprefix\url{https://link.aps.org/doi/10.1103/PhysRevLett.101.257003}.

\bibitem{BaFeAs-magmom2}
\bibinfo{author}{Rotter, M.} \emph{et~al.}
\newblock \bibinfo{title}{Competition of magnetism and superconductivity in
  underdoped \ce{(Ba_{1-x}K_x)Fe2As2}}.
\newblock \emph{\bibinfo{journal}{New J. Phys.}} \textbf{\bibinfo{volume}{11}},
  \bibinfo{pages}{025014} (\bibinfo{year}{2009}).
\newblock \urlprefix\url{https://doi.org/10.1088/1367-2630/11/2/025014}.

\end{thebibliography}
\end{document}